\documentclass[review]{elsarticle}

\usepackage{lineno,hyperref}
\usepackage{amsmath}
\usepackage{amssymb}
\usepackage{algorithm}
\usepackage[noend]{algpseudocode}
\usepackage{multicol}
\usepackage{caption}
\newcommand{\algrule}[1][.2pt]{\par\vskip.5\baselineskip\hrule height #1\par\vskip.5\baselineskip}

\modulolinenumbers[5]

\journal{Journal of \LaTeX\ Templates}

\newtheorem{thm}{Theorem}
\newtheorem{prop}[thm]{Proposition}
\newproof{pf}{Proof}
\newcommand\doubleplus{\ensuremath{\mathbin{+\mkern-10mu+}}} 

\begin{document}

\begin{frontmatter}

\title{BSF: a parallel computation model for scalability estimation of iterative numerical algorithms on cluster computing systems}

\author{Leonid B. Sokolinsky}
\address{South Ural State University (National Research University)
\\ 76, Lenin prospekt, Chelyabinsk, Russia, 454080
\\ E-mail: leonid.sokolinsky@susu.ru
\\ Phone: +79048040330}

\begin{abstract}
This paper examines a~novel parallel computation model called bulk synchronous farm (BSF) that focuses on estimating the scalability of compute-intensive iterative algorithms aimed at cluster computing systems. The main advantage of the proposed model is that it allows to estimate the scalability of a parallel algorithm before its implementation. Another important feature of the BSF model is the representation of problem data in the form of lists that greatly simplifies the logic of building applications. In the BSF model, a~computer is a~set of processor nodes connected by a~network and organized according to the master/slave paradigm. A cost metric of the BSF model is presented. This cost metric requires the algorithm to be represented in the form of operations on lists. This allows us to derive an equation that predicts the scalability boundary of a~parallel program: the maximum number of processor nodes after which the speedup begins to decrease. The paper includes examples of applying the BSF model to designing and analyzing parallel numerical algorithms. The large-scale computational experiments conducted on a~cluster computing system confirm the adequacy of the analytical estimations obtained using the BSF model.
\end{abstract}

\end{frontmatter}


\section{Introduction}\label{Section:Introduction}

Currently, we are entering the era of exascale computers operating at a~speed of 10 to the 18th power~$(10^{18})$ flops~\cite{1}. The recent TOP500 list (November 2020)~\cite{2} shows that more than 90\% of the most powerful supercomputers have the cluster architecture. The design of numerical algorithms for such computing clusters requires new approaches to achieve the high efficiency of parallelization. It is important to evaluate the scalability of a~parallel algorithm at an early stage of its development. \emph{Scalability} has been widely used in practice to describe how system sizes and problem sizes influence the performance of parallel computers and algorithms~\cite{3}. The main measure for evaluating the scalability of a~parallel algorithm on a~cluster computing system is the \emph{speedup}~$a$, which is defined as the ratio of the algorithm execution time~${T_1}$ on one processor node to the algorithm execution time~${T_K}$ on~$K$ processor nodes:
\begin{equation}\label{equation:01}a(K) = \frac{{{T_1}}}{{{T_K}}}.\end{equation}
It is well known that for a~given computing cluster architecture and a~fixed-size problem, the speedup of a~parallel algorithm does not continue to increase with an increase in the number of processor nodes, but it tends toward saturation and culminates in a~peak at a~certain system size after which the speedup begins to decrease. Let us define the \emph{scalability boundary} of the parallel algorithm as the number of processor nodes~${K_{\max }}$ at which the speedup peak is reached for the given problem size on the target cluster computing system. To detect the scalability boundary of a~ parallel algorithm, we have the following two possibilities. First, we can conduct a~series of large-scale computational experiments on the target cluster system to plot the speedup curve and visually determine the scalability boundary. However, it takes time and effort to build a~compilable and executable implementation of the parallel algorithm in some programming language. Moreover, we need to obtain access to a~sufficiently large cluster computing system for a~sufficiently long time. The second possibility is to use a~suitable parallel computation model that can predict the execution time of the algorithm for the target cluster computing system.

\begin{figure}[t]
  \centering
  \includegraphics[scale=0.9]{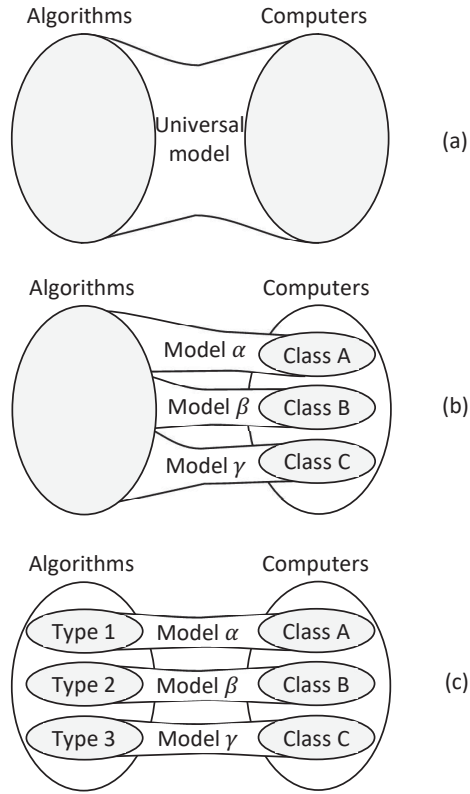}
  \caption{Models bridging algorithms and computers.}
  \label{Fig1}
\end{figure}

The \emph{computational model} is a~simplified and abstract description of a~computer. A computer architect, algorithm designer and program developer can use such a~model as a~basis to assess their work, including the suitability of one computer architecture to various applications, the computation complexity of an algorithm and the potential performance of one program on various computers, etc. A good computational model can simplify the complicated work of the architect, algorithm designer and program developer while mapping their work effectively onto real computers~\cite{4}. Thus, such a~ computational model is sometimes also called a~``bridging model''~\cite{5}. An \emph{universal} bridging model can be applied to any algorithms and any computers (see Fig.~\ref{Fig1}~a). The \emph{RAM} (\emph{random access machine}) \emph{model}~\cite{6,7,8} was such a universal model bridging the sequential computers and algorithms. With the advent of parallel computers, numerous attempts were made to build a~similar universal model bridging the multiprocessors and parallel algorithms~\cite{9}, but these attempts failed. This is mainly due to the large variety of multiprocessor architectures that are rapidly emerging and developing in response to the demands of increasing computer performance. Under these conditions, creating a~simple and accurate universal model of parallel computations is almost impossible. The approach schematically shown in Fig.~\ref{Fig1}~b was applied to overcome these difficulties~\cite{10}. According to this approach, the parallel architectures were divided into three classes: shared memory, distributed memory, and hierarchical memory multiprocessors~\cite{4}.

Separate parallel computation models were created for each class of multiprocessors, but almost all these models were universal with respect to a~variety of parallel numerical algorithms. This approach generated simple and reliable models with a~high level of abstraction, such as PRAM~\cite{11}, BSP~\cite{5}, and LogP~\cite{12}. Numerous attempts have been made to refine and extend these models to adapt them to the increasing complexity of multiprocessor system architectures. This has led to the emergence of more accurate but complicated to apply models of parallel computations (see, for example,~\cite{13,14,15,16,17,18}). Dividing the entire set of algorithms into different types allows us to correct this situation to a~certain extent (see Fig.~\ref{Fig1}~c). Examples of different types of algorithms can be iterative numerical algorithms, graph algorithms, big data processing algorithms, and so on. Each pair (algorithm type, architecture class) can have its own parallel computation model. Such an approach allows us to reach an acceptable tradeoff between the accuracy of estimations and usability.

G.~Bilardi and A.~Pietracaprina distinguish the following four typical components of a~computational model~\cite{19}: an \emph{architectural component}, described as an interconnection of modules of various functionalities; a~\emph{specification component}, determining what is a~(syntactically) valid algorithm/program; an \emph{execution component}, defining which sequences of states of the architectural modules constitute valid executions of a~program/algorithm on a~given input; and a~\emph{cost component}, defining one or more cost metrics for each execution. They also formulate the following three conflicting requirements by which a~computational model can be evaluated: the ease of algorithm/program design and analysis (\emph{usability}); the ability of estimating, from the cost metrics provided by the model, the actual performance of a~program on a~specific real platform (\emph{effectiveness}) and on a~class of hardware platforms (\emph{hardware} \emph{portability}). In accordance with the approach shown in Fig.~\ref{Fig1}~c, we add one more component: \emph{algorithmic portability}, determining the type of algorithms to which the model is applicable. In the context of this approach, this article presents a~novel model of parallel computations, named \emph{bulk synchronous farm} (\emph{BSF}) that is an extension of the BSP model proposed initially by Valiant in~\cite{5}. The BSF model is based on the original idea outlined in~\cite{20} and intended for evaluating the parallel iterative compute-intensive numerical algorithms on the cluster computing systems. Let us clarify the meaning of the restrictions imposed on the scope of the BSF model's applicability. The ``\emph{iterative algorithm}'' means that the time spent on the preparation to perform the iterative computations (data input, memory allocation, variable initialization, and so on) is so markedly less than the time spent on the iterative computations themselves that we can neglect the initialization overhead within the model. In addition, the framework of the ``\emph{iterative algorithm}'' includes a~sequence of steps implemented as a~loop in which the next iteration depends on the results of the previous ones and cannot be executed in parallel with them. The ``\emph{compute-intensive numerical algorithm}'' means that the time spent for calculations is greater or comparable to the time spent for input/output and communications between processor nodes. The ``\emph{cluster computing system}'' is a~set of tightly connected homogeneous processor nodes with private memory that communicate with each other through the MPI library. The model treats the processor node as a~black box that can perform scalar and vector operations at a~certain speed. The restrictions mentioned above allow us to obtain a~simple and reliable equation for estimating the scalability boundary of an iterative numerical algorithm by using the BSF model. No other known model provides such an equation.

The rest of the article is organized as follows.
Section~\ref{Section:Related_work} briefly reviews parallel computation models for distributed memory multiprocessors.
Section~\ref{Section:BSF_model} introduces the BSF model description determining its architectural, specification and execution components.
Section~\ref{Section:Cost_metric} presents the cost metric of the BSF model and contains the derivation of the main equation that estimates the scalability boundary of a~parallel algorithm. Section~\ref{Section:Jacobi_method} demonstrates how to apply the BSF model to estimate the algorithm scalability boundary using the iterative Jacobi method as an example. Section~\ref{Section:Experiments} includes the results of large-scale computational experiments and their comparison with the results obtained analytically using the BSF model. In Section~\ref{Section:Discussion}, we discuss the strengths and weaknesses of the BSF model. Section~\ref{Section:Conclusions} concludes this paper.

\section{Related works}\label{Section:Related_work}
One of the first parallel computation models for distributed memory multiprocessors was BSP (bulk-synchronous parallel) model proposed by Valiant in~\cite{5}. A \emph{BSP\nobreakdash-\hspace{0pt}computer} is a system of $K$ processors that have private memory and are connected by a network allowing data to be transferred from one processor to another. The following cost parameters of the interconnect are defined:
$g$~---~the time required to transfer a single machine word across the network;
$L$~---~the time required for a global synchronization.
In the BSP\nobreakdash-\hspace{0pt}computer, the message transfer is simulated using the notion of $h$\nobreakdash-\hspace{0pt}session. The \emph{h\nobreakdash-\hspace{0pt}session} is an abstraction of an arbitrary communication operation in which each processor transfers no more than $h$ machine words and receives no more than $h$ machine words. In the BSP\nobreakdash-\hspace{0pt}computer, the execution time of a single $h$\nobreakdash-\hspace{0pt}session cannot exceed $hg$. The \emph{BSP\nobreakdash-\hspace{0pt}program} consists of $n$ parallel processes each of that is assigned to a separate processor. The \emph{BSP\nobreakdash-\hspace{0pt}program} is divided into global sequential \emph{supersteps} that are synchronously executed by all processes. Each superstep includes the following four sequential steps: 1)~computations on each processor using only local data; 2)~global barrier synchronization; 3)~data transfer from any processor to any other processors by performing a single $h$\nobreakdash-\hspace{0pt}session; 4)~global barrier synchronization. In the BSP model, the \emph{cost metric} is constructed as follows. Let the BSP\nobreakdash-\hspace{0pt}program consist of $S$ supersteps. Assume that each processor performs no more than $w_i$ clock cycles during local calculations in the $i$\nobreakdash-\hspace{0pt}th superstep. The total time $t_i$ taken by the system to execute the $i$\nobreakdash-\hspace{0pt}th superstep is calculated using the equation \[{t_i} = {w_i}\; + \;hg + \;L.\] The total time $T$ of executing the entire program is determined by the equation \[T = W\; + \;hgS\; + \;LS,\] where $W = \sum\nolimits_{i = 1}^S {{w_i}}$. The BSP model has a number of disadvantages. The first, the BSP model admits to transfer messages with a length of only one machine word and does not take into account that transferring $h$ machine words as a single message can be more efficient than transferring $h$ messages with the length of one machine word. The second, the amount of data processed by a single processor in each superstep should be approximately equal to the number of words received during the $h$\nobreakdash-\hspace{0pt}session, i.e. $h$  that limits the granularity of parallelism from below. The third, the BSP model assumes hardware support for the global synchronization, but most multiprocessor systems with distributed memory do not have such mechanism.

To overcome these disadvantages, Culler and co-authors proposed a parallel computation model named LogP~\cite{12} that extends the BSP model. Like the BSP model, the \emph{LogP\nobreakdash-\hspace{0pt}computer} is a system of $P$ processors that have private memory and are connected by a network that allows data to be transferred from one processor to another.
The LogP model has the following cost parameters:
\begin{tabbing}
MM. \= M \= MMMMMMMMMMMMMMMMMMMMMMMMMMMMMMMMMMMMMMMMMMMMMMMMMMMMMMMMMMMMMMM \kill
$L$\> : \> \emph{latency} (time of transferring one machine word from one processor to\\
\>\> another);\\
$o$\> : \> \emph{overhead} (length of time that a processor is engaged in the transfer\\
\>\> or reception of each message);\\
$g$\> : \> \emph{gap} (minimum time interval between consecutive message transfers or\\
\>\> consecutive message receptions at a processor);\\
$P$\> : \> number of processors.
\end{tabbing}

The LogP model assumes that the network has a finite capacity, such that at most $\left\lceil {L/g} \right\rceil$ messages can be in transfer from any processor or to any processor at any time. Furthermore, the LogP model assumes that all messages have a small size (one or a few number of machine words). Large messages need to be fragmented. The time taken by computations using local data in the superstep is calculated in the same way as in the BSP model. The time of transferring one short message from one processor to another is $o + L + o$. The time of transferring $n$ consecutive short messages is $(n - 1)g + o + L + o$. An obvious disadvantage of the LogP model is the message size limitation.

The LogP model has been improved in numerous extensions. In~\cite{40}, an extension of the LogP model called LogGP was proposed. The \emph{LogGP model} adds the new parameter $G$ (gap per byte) that determines the time of transferring one byte within a long message. The time of transferring a message of length $m$ is $o + (m - 1)G + L + o$. The \emph{LogGPS model}~\cite{41} extends the LogGP model by introducing an additional parameter S that takes into account the synchronization overhead. Actually, the LogGPS model is an adaptation of the LogPQ model to the features of the communication protocols of the MPICH library~\cite{42} that is a portable implementation of MPI. Like the LogGPS model, the $log_nP$ \emph{model}~\cite{43} is dedicated to cluster computing systems that use MPI for message transfers. A distinctive feature of the $log_nP$ model is that it takes into account the implicit overhead of transferring data between different levels of hierarchical memory.

The emergence of heterogeneous cluster systems has generated a new class of parallel computation models. One of these models called \emph{Multi-BSP} was proposed by Valiant in~\cite{44}. The \emph{Multi-BSP model} extends the BSP model in the following two directions. First, Multi-BSP is a hierarchical model with an arbitrary number of the levels that represent the actual technical features of the hierarchical memory and various cache levels of modern multiprocessor systems. Second, Multi-BSP includes the amount of memory at each level as an additional parameter. For each level $i$ in the hierarchy, the following vector of parameters is introduced: $({p_i},{g_i},{L_i},{m_i})$. Here, ${p_i}$ is the number of processors, ${g_i}$ is a delay in transferring data from level $i$ to level $i+1$, ${L_i}$ is the synchronization overhead, and ${m_i}$ is the amount of memory/cache. For a system that includes $d$ levels, the total number of processors is calculated using the equation \mbox{${P_d} = \prod\nolimits_{i = 1}^d {{p_i}} $}, the total amount of memory is calculated using the equation \mbox{${M_d} = {m_d} + \sum\nolimits_{i = 1}^{d - 1} {{m_i}\prod\nolimits_{j = i + 1}^d {{p_j}} } $}, and the total delay is calculated using the equation \mbox{${G_d} = \sum\nolimits_{i = 1}^d {{g_i}} $}.

Another model of this class is the $mlog_nP$~\cite{45} extending the $log_nP$ model. The $mlog_nP$ \emph{model} is dedicated to computing clusters with multi-core processors. In this model, the $n$\nobreakdash-\hspace{0pt}level memory hierarchy introduced in the $log_nP$ model is called \emph{vertical}. Additionally, the authors introduce a \emph{horizontal} $m$\nobreakdash-\hspace{0pt}level hierarchy of data channels. The zero-level channel serves data exchange between the cores of a single processor, the first-level channel is used for data exchange between cores of different processors of a single processor node, the second-level channel is dedicated for data exchange between the cores of different processor nodes, and so on. The cost of transferring a message at $i$\nobreakdash-\hspace{0pt}th horizontal level is calculated using the equation \[{T_i} = \sum\limits_{j = 0}^{{n_i} - 1} {(o_j^i + l_j^i)}  = \sum\limits_{j = 0}^{{n_i} - 1} {(f_j^i(s,1) + f_j^i(s,d))} ,\] where the semantics of all parameters within a single horizontal level is inherited from the $log_nP$ model. The model is dedicated to predicting the execution time of a CUDA-program on cluster computing systems with Kepler GPUs.

In~\cite{37}, an extension of the BSP model for GPUs running CUDA is proposed. This model focuses on predicting the execution time of a CUDA-program on cluster computing systems with Kepler GPUs. The approximated execution time $T_k$ of a \emph{kernel function} with $t$ threads is calculated by the following equation \[{T_k} = \frac{{t \times (Comp + Com{m_{DRAM}} + Com{m_{GPUM}})}}{{R \times P \times \lambda }}.\] It sums the computational cost (\emph{Comp}) with the communication cost of global memory ($Comm_{GM}$) and shared memory ($Comm_{SM}$) accesses, performed by each thread. This cost is multiplied by the number of threads $t$ and divided by the clock rate $R$ times the number of cores $P$ available in the GPU. The parameter $\lambda$ is used to model the effects of application optimizations, such as divergence, shared bank conflicts and coalesced global memory accesses. The global communication cost is estimated in the same way as in the BSP model.

A large number of other sophisticated parallel computation models for modern cluster computing systems have been proposed in recent years (see surveys~\cite{4,10}). The main disadvantage of these models is the complexity of their practical application when designing and analyzing parallel numerical algorithms for exascale computers. None of these models yields a ready-to-use equation for estimating the scalability boundary of a parallel numerical algorithm.

\section{Description of BSF model}\label{Section:BSF_model}

\begin{figure}[t]
  \centering
  \includegraphics[scale=0.9]{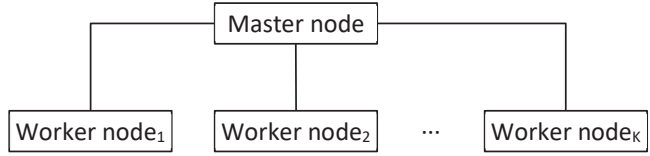}
  \caption{Architecture of BSF-computer.}
  \label{Fig2}
\end{figure}

In this section, we give a~description of the BSF model, namely, determine its architectural, specification and execution components. The \emph{architecture} of the BSF\nobreakdash-\hspace{0pt}computer is shown in Fig.~\ref{Fig2}. A BSF\nobreakdash-\hspace{0pt}computer consists of a~collection of homogeneous processor nodes with private memory connected by a~communication network delivering messages among the nodes. All the nodes have the same capacity. The BSF model treats the processor node of a~real computing cluster as a~black box. Node interaction is based on the master/slave paradigm~\cite{21}: one processor node is the \emph{master node}; all other processor nodes are \emph{worker nodes} (also sometimes referred to as ``slaves''). The master node serves as the control and communication hub.

In the BSF model, an algorithm must be \emph{specified} in the form of operations on lists by using the \emph{Map} and \emph{Reduce} functions. The higher-order functions \emph{Map} and \emph{Reduce}  defined in the Bird--Meertens formalism~\cite{22} are the basis for the parallelization of the BSF\nobreakdash-\hspace{0pt}algorithms. Let~$[{a_1}, \ldots ,{a_l}]$ denote a~list of length~$l$ that includes elements of a~given set~$\mathbb{A}$. Let~$F:\mathbb{A} \to \mathbb{B}$ be a~function that maps the set~$\mathbb{A}$ to a~given set~$\mathbb{B}$. The higher-order function \emph{Map} applies the function~$F$ to each element of the list~$[{a_1}, \ldots ,{a_l}]$ and returns a~list of results in the same order:
\begin{equation}\label{equation:02}Map\left( {F,\left[ {{a_1}, \ldots ,{a_l}} \right]} \right) = \left[ {F\left( {{a_1}} \right), \ldots ,F\left( {{a_l}} \right)} \right].\end{equation}
Let~$[{b_1}, \ldots ,{b_l}]$ denote a~list of length~$l$ that includes elements of the set~$\mathbb{B}$. Let~$\oplus :\mathbb{B}\times\mathbb{B}\to\mathbb{B}$ be a~binary associative operation on the set~$\mathbb{B}$. The higher-order function \emph{Reduce} reduces the list~$[{b_1}, \ldots ,{b_l}]$ to a~single value by iteratively applying the operation~$\oplus$ to its elements:
\begin{equation}\label{equation:03}Reduce( \oplus ,[{b_1}, \ldots ,{b_l}]) = {b_1} \oplus  \ldots  \oplus {b_l}.\end{equation}

\begin{algorithm}[t]
\caption{Generic BSF-algorithm template.}\label{alg1}
\begin{algorithmic}[1]
\State \textbf{input} $A,x^{(0)}$
\State $i := 0$
\State \label{alg1:loop_begin}$B := Map(F_{x^{(i)}},A)$
\State $s := Reduce(\oplus,B)$
\State $x^{(i+1)} := Compute(x^{(i)},s)$
\State $i := i+1$
\State \textbf{if} $StopCond(x^{(i)},x^{(i+1)})$ \textbf{goto} \ref{alg1:output}
\State \textbf{goto} \ref{alg1:loop_begin}
\State \label{alg1:output}\textbf{output} $x^{(i)}$
\State \textbf{stop}
\end{algorithmic}
\end{algorithm}

The generic template of an iterative BSF\nobreakdash-\hspace{0pt}algorithm is presented as Algorithm~\ref{alg1}. The variable~$i$ denotes the iteration number; ${x^{(0)}}$ is an initial approximation; ${x^{(i)}}$ is the \emph{i}th approximation (the approximation can be a~number, a~vector, or any other data structure); $A$ is the list of elements of a~certain set~$\mathbb{A}$, which represents the source data of the problem; ${F_x}:\mathbb{A} \to \mathbb{B}$ is a~parameterized function (the parameter~$x$ is the current approximation) that maps the set~$\mathbb{A}$ to a~certain set~$\mathbb{B}$; $B$~is a~list of elements of the set~$\mathbb{B}$ calculated by applying the function~${F_x}$ to each element of the list~$A$; $\oplus$ is a~binary associative operation on the set~$\mathbb{B}$. Step~1 reads the input data of the problem and the initial approximation. Step~2 assigns the zero value to the iteration counter~$i$. Step~3 calculates the list~$B$ by invocating the higher-order function~$Map({F_{{x^{(i)}}}},A)$. Step~4 assigns the result of the higher-order function~$Redice( \oplus ,B)$ to the intermediate variable~$s$. Step~5 invocates the user function \emph{Compute} that calculates the next approximation~${x^{(i + 1)}}$ taking two parameters: the current approximation~${x^{(i)}}$ and the result~$s$ of the higher-order function \emph{Reduce}. Step~6 increases the iteration counter~$i$ by one. Step~7 checks termination criteria by invocation of the user Boolean function \emph{StopCond}, which takes two parameters: the new approximation~${x^{(i)}}$ and the previous approximation~${x^{(i - 1)}}$. If \emph{StopCond} returns true, the algorithm outputs~${x^{(i)}}$ as an approximate problem solution and stops working. Otherwise, the control is passed to Step~3 starting the next iteration.

\begin{figure}[t]
  \centering
  \includegraphics[scale=0.9]{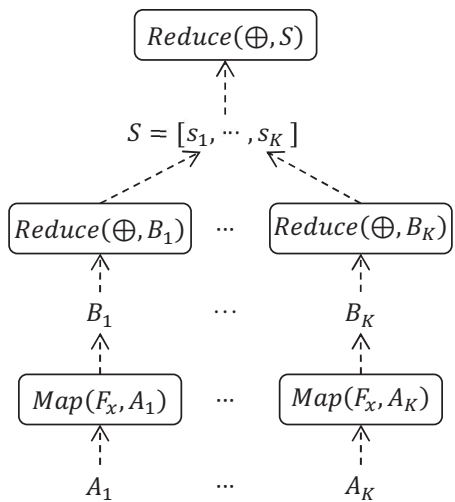}
  \caption{BSF-algorithm parallelization schema.}
  \label{Fig3}
\end{figure}

The \emph{parallel execution} of a~BSF\nobreakdash-\hspace{0pt}algorithm is based on the following theoretical foundation. Let us divide the list~$A = [{a_1}, \ldots ,{a_l}]$ into~$K$ sublists of length~$m$:
\begin{equation}\label{equation:04}A = A_1 \doubleplus \cdots \doubleplus A_K\end{equation}
(for simplicity, we assume that~$l$ is a~multiple of~$K$, i.e., $l = Km$ for some~$m \in \mathbb{N}$). Here,~$\doubleplus$~denotes the operation of list concatenation. According to the promotion theorem~\cite{23}, the following equation holds:
\begin{equation}\begin{gathered}\label{equation:05}
Reduce\left( { \oplus ,Map\left( {{F_x},A} \right)} \right) = \hfill \\
= Reduce\left( { \oplus ,Map\left( {{F_x},{A_1}} \right)} \right) \oplus  \cdots  \oplus Reduce\left( { \oplus ,Map\left( {{F_x},{A_K}} \right)} \right).
\end{gathered}\end{equation}
Equation~\eqref{equation:05} gives us the generic parallelization scheme shown in Fig.~\ref{Fig3}. We can run~$K$ parallel \emph{worker} threads that independently perform the higher-order functions \emph{Map} and \emph{Reduce} over sublists~${A_1}, \ldots ,{A_K}$. Then, the \emph{master} thread joins the produced partial foldings~${s_1}, \ldots ,{s_K}$ into the single list~$S$ and performs higher-order function \emph{Reduce} over it.

The generic parallelization template of an iterative BSF\nobreakdash-\hspace{0pt}algorithm is presented as Algorithm~\ref{alg2}. It includes~$K + 1$ parallel processes: one master process and~$K$ worker processes. The master process runs on the master node. Each worker process runs on a~separate worker node. In Step~1, the master process reads the initial approximation $x^{(0)}$ and assigns the zero value to the iteration counter $i$. At the same time, every \emph{j}th worker process reads the sublist $A_j$ that assigned to it for processing and is treated as local data. In Step~2, the master process sends the current approximation~${x^{(i)}}$ to all worker processes. After that, every \emph{j}th worker process independently applies higher-order functions \emph{Map} and \emph{Reduce} to its sublist (Steps 3 and 4). In Steps 3 and 4, the master process is idle. In Step~5, every \emph{j}th worker process sends to the master process the partial folding~${s_j}$ that is a result of the \emph{Reduce} function. In Steps 6-9, the master process performs the following actions: executes the higher-order function \emph{Reduce} over the list of partial foldings~$\left[ {{s_1}, \ldots ,{s_K}} \right]$; invocates the user function \emph{Compute} that calculates the next approximation; and checks the termination criteria by using the user Boolean function \emph{StopCond} and assigns its result to the Boolean variable \emph{exit}. In Steps 6-9, the worker processes are idle. In Step~10, the master process sends the \emph{exit} value to all worker processes. If the \emph{exit} value is false, the master process and worker processes go to the next iteration; otherwise, the master processes output the result and the computation stops. Note that in Steps 2 and 10, all processes perform the implicit global synchronization. In this template, all worker processes execute the same code for different sublists. Since all sublists have the same length, there is no need to balance the workload of the worker nodes.
\begin{algorithm}[t]
\caption{Generic BSF-algorithm parallelization template.}\label{alg2}
\begin{multicols}{2}
\begin{center}
\textbf{Master} \\
\textbf{\emph{j}th Worker (j=1,\dots,K)}
\end{center}
\end{multicols}
\algrule
\begin{multicols}{2}
\begin{algorithmic}[1]
\State \textbf{input} $x^{(0)}; i := 0$
\State\label{alg2:Master_loop_begin}$SendToAllWorkers(x^{(i)})$
\State
\State
\State $RecvFromWorkers\left(s_1,\ldots,s_K\right)$
\State $s:= Reduce\left(\oplus,[s_1,\dots,s_K]\right)$
\State $x^{(i+1)} := Compute\left(x^{(i)},s\right)$
\State $i := i+1$
\State $exit := StopCond\left(x^{(i)},x^{(i+1)}\right)$
\State $SendToAllWorkers(exit)$
\State \textbf{if not} $exit$ \textbf{goto} \ref{alg2:Master_loop_begin}
\State \textbf{output} $x^{(i)}$
\State \textbf{stop}
\end{algorithmic}
\begin{algorithmic}[1]
\State \textbf{input} $A_j$
\State \label{alg2:Worker_loop_begin}$RecvFromMaster\left(x^{(i)}\right)$
\State $B_j := Map(F_{x^{(i)}},A_j)$
\State $s_j := Reduce(\oplus,B_j)$
\State $SendToMaster(s_j)$
\State
\State
\State
\State
\State $RecvFromMaster(exit)$
\State \textbf{if not} $exit$ \textbf{goto} \ref{alg2:Worker_loop_begin}
\State
\State \textbf{stop}
\end{algorithmic}
\end{multicols}
\end{algorithm}

\section{Cost metric of BSF model}\label{Section:Cost_metric}

The BSF model assumes that the overhead of initializing and terminating a~program is negligible compared to the overhead of executing the iterative process. The cost of an iterative process is the sum of the costs of individual iterations. Therefore, to estimate the execution time of an iterative BSF\nobreakdash-\hspace{0pt}algorithm, we simply need to obtain an estimation of the time cost of one iteration. The BSF model includes the following cost parameters for a~single iteration:
\begin{tabbing}
MM. \= M \= MMMMMMMMMMMMMMMMMMMMMMMMMMMMMMMMMMMMMMMMMMMMMMMMMMMMMMMMMMMMMMM \kill
$K$\> : \> number of worker nodes;\\
$l$\> : \> length of the list $A$ representing the input data (the same as\\
\>\> the length of the list $B$ representing the result of the higher-order\\
\>\> function \emph{Map});\\
$L$\> : \> latency (time of transferring one-byte message node-to-node);\\
${t_c}$\> : \> time taken by the master node to send the current approximation to\\
\>\> and receive a folding from one worker node (including latency);\\
${t_{Map}}$\> : \> time taken by a~single worker node to execute the higher-order\\
\>\> function \emph{Map} over the entire list $A$;\\
${t_{Rdc}}$\> : \> time taken by a~single worker node to execute the higher-order\\
\>\> function \emph{Reduce} over the entire list $B$;\\
${t_p}$\> : \> time taken by the master node to process the result received from\\
\>\> the worker nodes and check the termination criteria (steps 7 and 9 that\\
\>\> do not depend on $K$).
\end{tabbing}
We will also use the parameter $t_a$ that denotes the time taken by a~node (master or worker) to execute the operation $\oplus$ being the second parameter of the higher-order function \emph{Reduce}:
\begin{equation}\label{equation:06}
t_a = \frac{t_{Rdc}}{l-1}.
\end{equation}

\begin{figure}[t]
  \centering
  \includegraphics[scale=0.9]{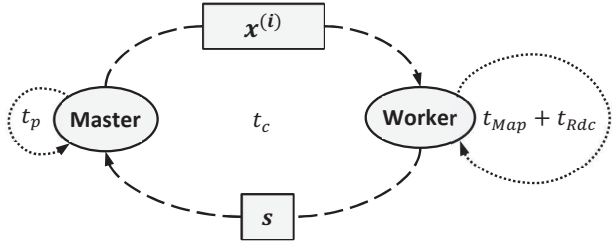}
  \caption{Diagram of Algorithm~2 for the configuration with one master and one worker.}
  \label{Fig4}
\end{figure}

First, let us consider the performance of Algorithm~2 on a~BSF\nobreakdash-\hspace{0pt}computer consisting of one master node and one worker node (see Fig. \ref{Fig4}). Here, the dashed arrows denote the data transfers, and the dotted arrows denote the computation loops. Let~${T_1}$ denote the execution time of one iteration of Algorithm~2 by a~BSF\nobreakdash-\hspace{0pt}computer with one master node and one worker node. Using the cost parameters introduced above, we obtain the following estimation of the time~${T_1}$:
\begin{equation}\label{equation:07}
{T_1} = {t_p} + {t_c} + t_{Map} + t_{Rdc}.
\end{equation}

\begin{figure}[t]
  \centering
  \includegraphics[scale=0.9]{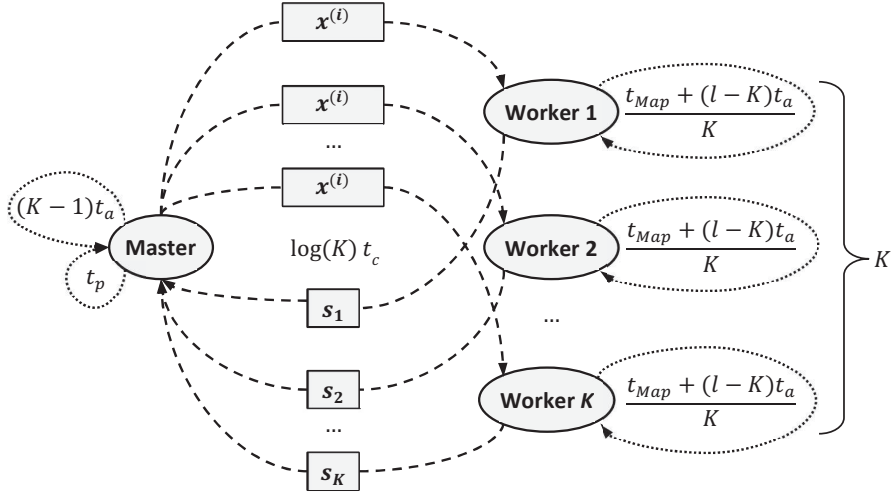}
  \caption{Diagram of Algorithm~2 for the configuration with one master and~$K$ workers.}
  \label{Fig5}
\end{figure}

Second, let us consider the performance of Algorithm~2 on a~BSF\nobreakdash-\hspace{0pt}computer consisting of one master node and~$K$ worker nodes (see Fig. \ref{Fig5}). We assume from now on that $l\geq K$. Let~${T_K}$ denote the execution time of one iteration of Algorithm~2 by a~BSF\nobreakdash-\hspace{0pt}computer with one master node and~$K$ worker nodes. It is known that a good MPI implementation would implement a broadcast or allreduce for $K$ processes with $O(\log K)$~\cite{36a}. If we use \emph{MPI\_Broadcast} to implement Step 2 and \emph{MPI\_Reduce} to implement Step 5 of Algorithm~\ref{alg2} then we can obtain the following estimation for~$T_K$:
\begin{equation}\label{equation:08}
{T_K} = (K - 1){t_a} + {t_p} + \left( {{{\log }_2}\left( K \right) + 1} \right){t_c} + \frac{{{t_{Map}} + (l - K){t_a}}}{K}.
\end{equation}
Note that for $K=1$, this equation is converted to equation~\eqref{equation:07}.

In the BSF model, the speedup~$a$ as a~function of~$K$ is calculated as follows:
\begin{equation}\label{equation:09}
a_{BSF}(K) = \frac{{{T_1}}}{{{T_K}}} = \frac{{{t_p}{\text{ }} + {\text{ }}{t_c}{\text{ }} + {\text{ }}{t_{Map}}{\text{ }} + {\text{ }}{t_{Rdc}}}}{{(K - 1){t_a} + {t_p} + \left( {{{\log }_2}\left( K \right) + 1} \right){t_c} + \frac{{{t_{Map}} + (l - K){t_a}}}{K}}},
\end{equation}
where~$K$ is the number of worker nodes. For positive values of all parameters and for~$K \geqslant 1$, the function~$a_{BSF}(K)$ defined by equation~\eqref{equation:09} has the following properties:
\begin{equation}\label{equation:10}a_{BSF}(1) = 1;\end{equation}
\begin{equation}\label{equation:11}a_{BSF}(K) > 0;\end{equation}
\begin{equation}\label{equation:12}\mathop {\lim }\limits_{t_{comp} \to 0} a_{BSF}(K) = \frac{1}{\log_2 \left( K \right) + 1},\end{equation}
where~$t_{comp} = t_{Map} + t_{Rdc} + {t_p}$. All of these properties follow directly from equation~\eqref{equation:09} and do not require proofs. From the content point of view, property~\eqref{equation:10} corresponds to reality: speedup on a~single worker node must be equal to 1. Property~\eqref{equation:11} also confirms the adequacy of equation~\eqref{equation:09} since the speedup is always a~positive quantity. Property~\eqref{equation:12} says that for very small values of parameters~$t_{Map}$,~$t_{Rdc}$ and~$t_p$ that determine the total time of computations, equation~\eqref{equation:09} tends to equation~$a_{BSF}(K) = 1/\left( \log_2 \left( K \right) + 1\right)$. The last one, on the interval~$[1, + \infty )$, determines a~monotonically decreasing function that has a~maximum value~$1$ for~$K = 1$. This means that the BSF model is not applicable for algorithms in which the time spent on data transferring between the processor nodes is incomparably greater than the time spent on computations. In this case, one should use another parallel computation model (see the survey~\cite{10}). We state the main property of equation~\eqref{equation:09} in the form of the following proposition:

\begin{prop}\label{Proposition1}
Let $l \in \mathbb{N}$; $L,{t_c},{t_p} \in {\mathbb{R}_{ > 0}}$; ${t_{Map}},{t_a} \in {\mathbb{R}_{ \geqslant 0}}$; ${t_{Map}} + {t_a} > 0$. Then, the function $a_{BSF}(K)$ defined by equation~\eqref{equation:09} has a~single extremum on the interval~$(1, + \infty )$, which is the maximum.
\end{prop}
\begin{pf}
To find the extrema of function~\eqref{equation:09}, let us calculate the derivative of the speedup with respect to~$K$:
\[ a'(K) =\frac{{\left( {{t_p} + {t_c} + {t_{Map}} + (l - 1){t_a}} \right) \cdot \left( {\frac{{{t_a}K + {t_{Map}} + (l - K){t_a}}}{{{K^2}}} - {t_a} - \frac{{{t_c}}}{{K\ln 2}}} \right)}}{{{{\left( {(K - 1){t_a} + {t_p} + \left( {{{\log }_2}\left( K \right) + 1} \right){t_c} + \frac{{{t_{Map}} + (l - K){t_a}}}{K}} \right)}^2}}}.\]
Multiplying the numerator and denominator by $K^2$ and combining the like terms in the denominator, we obtain
{\small\begin{equation}\label{equation:13}
a'(K) = \frac{{\left( {{t_p} + {t_c} + {t_{Map}} + (l - 1){t_a}} \right) \cdot \left( { - {t_a}{K^2} - K{t_c}/\ln 2 + {t_{Map}} + l{t_a}} \right)}}{{{{\left( {K(K - 1){t_a} + K{t_p} + K\left( {{{\log }_2}\left( K \right) + 1} \right){t_c} + {t_{Map}} + (l - K){t_a}} \right)}^2}}}.
\end{equation}}
Extrema are reached at points where the derivative is zero. Therefore, we need to solve the following equation:
\[\frac{{\left( {{t_p} + {t_c} + {t_{Map}} + (l - 1){t_a}} \right) \cdot \left( { - {t_a}{K^2} - K{t_c}/\ln 2 + {t_{Map}} + l{t_a}} \right)}}{{{{\left( {K(K - 1){t_a} + K{t_p} + K\left( {{{\log }_2}\left( K \right) + 1} \right){t_c} + {t_{Map}} + (l - K){t_a}} \right)}^2}}}=0.\]
Under the conditions of the proposition, the first factor of numerator and the denominator in this equation are positive for all $K>0$. Hence, this equation is equivalent to the following quadratic equation
\[ - {t_a}{K^2} - ({t_c}/\ln 2 + {t_a})K + {t_{Map}} + l{t_a}=0\]
that has only one root on the interval $[1, + \infty )$:
\[
K_0 =\frac{1}{2}\sqrt {{{\left( {\frac{{{t_c}}}{{{t_a}\ln 2}}} \right)}^2} + \frac{{{t_{Map}}}}{{{t_a}}} + 4l}  - \frac{{{t_c}}}{{{t_a}\ln 2}}.
\]
Since $K^2$ has a negative coefficient, the derivative~$a'(K)$ calculated by equation~\eqref{equation:13} takes only positive values in the interval~$[1,{K_0})$ and only negative values in the interval~$({K_0}, + \infty )$. Therefore, the point~${K_0}$ is the maximum of the function~$a_{BSF}(K)$ on the interval~$[1, + \infty )$. The proposition is proven.
\end{pf}

Proposition \ref{Proposition1} gives us the following equation to evaluate the scalability boundary of a~BSF\nobreakdash-\hspace{0pt}algorithm:
\begin{equation}\label{equation:14}
{K_{BSF}} = \frac{1}{2}\sqrt {{{\left( {\frac{{{t_c}}}{{{t_a}\ln 2}}} \right)}^2} + \frac{{{t_{Map}}}}{{{t_a}}} + 4l}  - \frac{{{t_c}}}{{{t_a}\ln 2}}.
\end{equation}
It is noteworthy that the scalability boundary of a parallel BSF-algorithm does not depend on the time $t_p$ taken by the master node to process the result received from the worker nodes and check the termination criteria (steps~7 and~9 of Algorithm~\ref{alg2}). This is quite natural. Indeed, $t_p$~does not depend on the number of workers~$K$ and therefore cannot affect the point of speedup curve maximum that is completely determined by the derivative of the speedup with respect to~$K$.

\section{Applying the BSF model to the Jacobi method}\label{Section:Jacobi_method}

In this section, we will show how to apply the BSF model to estimate the scalability boundary of a parallel algorithm without software implementation and computational experiments. As an example, we use the Jacobi iterative method. The \emph{Jacobi method}~\cite{24} is a~simple iterative method for solving a~system of linear equations. This method was originally described by the German mathematician Carl Gustav Jacob Jacobi in~\cite{25}. Let us give a~brief description of the Jacobi method.

Let a~joint square system of linear equations in a~matrix form be given in Euclidean space~${\mathbb{R}^n}$:
\begin{equation}\label{equation:15}Ax = b,\end{equation}
where
\[\begin{gathered}
  A = \left( {\begin{array}{*{20}{c}}
  {{a_{11}}}& \cdots &{{a_{1n}}} \\
   \vdots & \ddots & \vdots  \\
  {{a_{n1}}}& \cdots &{{a_{nn}}}
\end{array}} \right); \hfill \\
  x = ({x_1}, \ldots ,{x_n}); \hfill \\
  b = ({b_1}, \ldots ,{b_n}). \hfill \\
\end{gathered} \]
It is assumed that~${a_{ii}} \ne 0$ for all~$i = 1, \ldots ,n$. Let us define the matrix
\[C = \left( {\begin{array}{*{20}{c}}
  {{c_{11}}}& \cdots &{{c_{1n}}} \\
   \vdots & \ddots & \vdots  \\
  {{c_{n1}}}& \cdots &{{c_{nn}}}
\end{array}} \right)\]
in the following way:
\[{c_{ij}} = \left\{ {\begin{array}{*{20}{l}}
  { - \frac{{{a_{ij}}}}{{{a_{ii}}}},\forall j \ne i;} \\
  {0,\forall j = i.}
\end{array}} \right.\]
Let us define the vector $d = ({d_1}, \ldots ,{d_n})$ as follows: ${d_i} = {{{b_i}} \mathord{\left/
 {\vphantom {{{b_i}} {{a_{ii}}}}} \right.
 \kern-\nulldelimiterspace} {{a_{ii}}}}$. The Jacobi method of finding an approximate solution of system~\eqref{equation:15} consists of the following steps:
\begin{enumerate}[Step 1.]
\item $k: = 0$; ${x^{(0)}}: = d$.\label{JacobiStep1}
\item ${x^{(k + 1)}}: = C{x^{(k)}} + d$.\label{JacobiStep2}
\item If ${\left\| {{x^{(k + 1)}} - {x^{(k)}}} \right\|^2} < \varepsilon $, go to Step~5.\label{JacobiStep3}
\item $k: = k + 1$; go to Step~2.\label{JacobiStep4}
\item Stop.\label{JacobiStep5}
\end{enumerate}
In the Jacobi method, an arbitrary vector~${x^{(0)}}$ can be taken as the initial approximation. In Step~\ref{JacobiStep1}, the initial approximation~${x^{(0)}}$ is assigned by the vector~$d$. In Step~\ref{JacobiStep3}, the Euclidean norm~$\left\|  \cdot  \right\|$ is used in the termination criteria. The \emph{diagonal dominance} of the matrix~$A$ is a~sufficient condition for the convergence of the Jacobi method:
\[\left| {{a_{ii}}} \right| \geqslant \left( {\sum\limits_{j = 1}^n {\left| {{a_{ij}}} \right|} } \right) - \left| {{a_{ii}}} \right|\]
for all~$i = 1, \ldots ,n$, and at least one inequality is strict. In this case, the system~\eqref{equation:15} has a~unique solution for any right-hand side.

Let us represent the Jacobi method in the form of an algorithm on lists. Let~${c_j}$ denote the~$j$\nobreakdash-\hspace{0pt}th column of matrix~$C$:
\[{c_j} = \left( {\begin{array}{*{20}{c}}
  {{c_{1j}}} \\
   \vdots  \\
  {{c_{nj}}}
\end{array}} \right).\]
Let $G = \left[ {1, \ldots ,n} \right]$ be the list of natural numbers from~$1$ to~$n$. For any vector $x = ({x_1}, \ldots ,{x_n}) \in {\mathbb{R}^n}$, let us define the function ${F_x}:\left\{ {1, \ldots ,n} \right\} \to {\mathbb{R}^n}$ as follows:
\begin{equation}\label{equation:16}{F_x}(j) = {x_j}{c_j} = \left( {\begin{array}{*{20}{c}}
  {{x_j}{c_{1j}}} \\
   \vdots  \\
  {{x_j}{c_{nj}}}
\end{array}} \right),\end{equation}
i.e., the function~${F_x}(j)$ multiplies the~$j$\nobreakdash-\hspace{0pt}th column of the matrix~$C$ by the $j$\nobreakdash-\hspace{0pt}th coordinate of the vector~$x$. The BSF implementation of the Jacobi method presented as Algorithm~\ref{alg3} can be easily obtained from the generic BSF\nobreakdash-\hspace{0pt}algorithm template (Algorithm~\ref{alg1}). In Algorithm~3,~$\vec  + $ and~$\vec  - $ denote the operations of vector addition and subtraction, respectively. Note that the matrix~$C$ entered in line~1 is implicitly used to calculate the values of the function~${F_{{x^{(k)}}}}$ in line 3.

\begin{algorithm}[t]
\caption{BSF-Jacobi algorithm.}\label{alg3}
\begin{algorithmic}[1]
\State \textbf{input} $C,G,d$
\State $k := 0; x^{(0)} := d$
\State \label{alg3:loop_begin}$B := Map(F_{x^{(k)}},G)$
\State $s := Reduce(\vec +,B)$
\State $x^{(k+1)} := s \vec + d$
\State $k := k+1$
\State \textbf{if} ${\left\| {{x^{(k + 1)}} \vec - {x^{(k)}}} \right\|^2} < \varepsilon$ \textbf{goto} \ref{alg3:output}
\State \textbf{goto} \ref{alg3:loop_begin}
\State \label{alg3:output}\textbf{output} $x^{(k)}$
\State \textbf{stop}
\end{algorithmic}
\end{algorithm}

The \emph{BSF-Jacobi parallel algorithm} (see Algorithm~\ref{alg4}) is automatically generated from Algorithm~3 by using the generic BSF\nobreakdash-\hspace{0pt}algorithm parallelization template (Algorithm~\ref{alg2}). Let us evaluate this parallel algorithm by using the BSF model. We assume that all arithmetic operations (addition and multiplication) as~well as the comparison operation of floating-point numbers take the same time, which we denote as~${\tau _{op}}$. To perform the scalability analysis of the BSF Jacobi algorithm, let us introduce the following notation (all quantities are taken with respect to a~ single iteration):
\begin{algorithm}[t]
\caption{BSF-Jacobi parallel algorithm.}\label{alg4}
\begin{multicols}{2}
\begin{center}
\textbf{Master} \\
\textbf{\emph{j}th Worker (\emph{j}=1,\dots,\emph{K})}
\end{center}
\end{multicols}
\algrule
\begin{multicols}{2}
\begin{algorithmic}[1]
\State \textbf{input} $d$
\State $k := 0; x^{(0)} := d$
\State\label{alg4:Master_loop_begin}$SendToAllWorkers(x^{(k)})$
\State
\State
\State $RecvFromWorkers\left(s_1,\ldots,s_K\right)$
\State $s:= Reduce\left(\vec +,[s_1,\dots,s_K]\right)$
\State $x^{(k+1)} := s\vec + d$
\State $k := k+1$
\State $exit := {\left\| {{x^{(k + 1)}} \vec - {x^{(k)}}} \right\|^2} < \varepsilon$
\State $SendToAllWorkers(exit)$
\State \textbf{if not} $exit$ \textbf{goto} \ref{alg4:Master_loop_begin}
\State \textbf{output} $x^{(k)}$
\State \textbf{stop}
\end{algorithmic}
\begin{algorithmic}[1]
\State \textbf{input} $C_j,G_j$
\State
\State \label{alg4:Worker_loop_begin}$RecvFromMaster\left(x^{(k)}\right)$
\State $B_j := Map(F_{x^{(k)}},G_j)$
\State $s_j := Reduce(\vec +,B_j)$
\State $SendToMaster(s_j)$
\State
\State
\State
\State
\State $RecvFromMaster(exit)$
\State \textbf{if not} $exit$ \textbf{goto} \ref{alg4:Worker_loop_begin}
\State
\State \textbf{stop}
\end{algorithmic}
\end{multicols}
\end{algorithm}
\begin{tabbing}
MM. \= M \= MMMMMMMMMMMMMMMMMMMMMMMMMMMMMMMMMMMMMMMMMMMMMMMMMMMMMMMMMMMMMMM \kill
${c_c}$\> : \> the quantity of real numbers that the master sends to and receives\\
\>\>  from a~single worker within one iteration;\\
$c_{Map}$\> : \> the quantity of arithmetic operations performed in Step~3 of the\\
\>\> Algorithm~\ref{alg3};\\
$c_a$\> : \> the quantity of arithmetic operations required to calculate the sum\\
\>\> of two vectors.\\
\end{tabbing}
Let us calculate these quantities. At the beginning of the iteration, the master sends to each worker the current approximation~${x^{(k)}}$, which is a~vector of length~$n$. At the ending of the iteration, each worker sends the calculated vector~$s_j$ of length $n$ to the master. Hence,
\begin{equation}\label{equation:17}{c_c} = 2n.\end{equation}
The higher-order function $Map\left( {{F_{{x^{(k)}}}},G} \right)$, in this case, multiplies all columns of the matrix~$C$ by the corresponding coordinates of the vector~$x$. Consequently,
\begin{equation}\label{equation:18}{c_{Map}} = {n^2}.\end{equation}
Adding two vectors of length~$n$ requires~$n$ arithmetic operations. Thus,
\begin{equation}\label{equation:19}c_a = n.\end{equation}

Let~${\tau _{op}}$ be the average execution time of a~single arithmetic or comparison operation by the processor node, and~${\tau _{tr}}$ be the average time for transferring a~single floating number across the network excluding latency. Using~\eqref{equation:17}-\eqref{equation:19}, we obtain the following values of the cost parameters of the BSF\nobreakdash-\hspace{0pt}Jacobi parallel algorithm:
\begin{equation}\label{equation:20}{t_c} = c_c\tau _{tr}+2L = 2(n\tau _{tr}+L);\end{equation}
\begin{equation}\label{equation:21}{t_{Map}} = {c_{Map}}{\tau _{op}} = n^2\tau _{op};\end{equation}
\begin{equation}\label{equation:22}t_a = c_a\tau _{op} = n\tau _{op}.\end{equation}
For the BSF\nobreakdash-\hspace{0pt}Jacobi algorithm, the length $l$ of the list is equal to the space dimension $n$:
\begin{equation}\label{equation:23}l=n.\end{equation}
Substituting the values of the right-hand sides of equations~\eqref{equation:20}-\eqref{equation:23} into equation~\eqref{equation:14}, we obtain the following equation for estimating the scalability boundary of the BSF\nobreakdash-\hspace{0pt}Jacobi parallel algorithm:
\begin{equation}\label{equation:24}
{K_{BSF - Jacobi}} = \sqrt {{{\left( {\frac{{n{\tau _{tr}} + L}}{{n{\tau _{op}}\ln 2}}} \right)}^2} + \frac{5}{2}n}  - \frac{{n{\tau _{tr}} + L}}{{n{\tau _{op}}\ln 2}}.
\end{equation}
For large values of~$n$, this is equivalent to
\begin{equation}\label{equation:25}{K_{BSF - Jacobi}} \approx O(\sqrt{n}) .\end{equation}
Therefore, we can conclude that the scalability boundary of the BSF\nobreakdash-\hspace{0pt}Jacobi parallel algorithm grows in proportion to the square root of the problem dimension~$n$. It should be noted that this result was obtained before a software implementation of the BSF-Jacobi parallel algorithm. In the next section, we verify this analytical estimation by computational experiments on a real cluster system.

\section{Computational experiments}\label{Section:Experiments}

For the rapid development of the parallel BSF\nobreakdash-\hspace{0pt}programs, the author implemented an algorithmic skeleton in C++ using the MPI parallel programming library~\cite{26}. The source code of this \emph{BSF\nobreakdash-\hspace{0pt}skeleton}~\cite{27} is freely available on GitHub, at \url{https://github.com/leonid-sokolinsky/BSF-skeleton}. Using the BSF\nobreakdash-\hspace{0pt}skeleton, we developed the parallel implementations of several iterative numerical methods and performed the computational experiments on the ``Tornado SUSU'' computing cluster~\cite{28}, whose specifications are shown in Table~\ref{Table1}. In this section, we present some of the results of these computational experiments and compare them with analytical results obtained by using the BSF cost metric.

\begin{table}[t]
\caption{Specifications of ``Tornado SUSU'' computing cluster.}
\centering
\begin{tabular}{l|l}
  \hline
  Parameter & 480 \\
  \hline
  Processor & Intel Xeon X5680 (6 cores, 3.33 GHz) \\
  Processors per node & 2\\
  Memory per node & 24 GB DDR3\\
  Interconnect & InfiniBand QDR (40 Gbit/s) \\
  Operating system & Linux CentOS\\
  \hline
\end{tabular}\label{Table1}
\end{table}

The first series of experiments was performed with the BSF\nobreakdash-\hspace{0pt}Jacobi parallel algorithm discussed in Section~\ref{Section:Jacobi_method}. The source code implemented by using the BSF\nobreakdash-\hspace{0pt}skeleton is freely available on GitHub, at \url{https://github.com/leonid-sokolinsky/BSF-Jacobi}. To carry out the experiments, we used a scalable system of linear equations~\eqref{equation:15} having the following coefficient matrix~$A$ and the vector of constant terms~$b$:
\[A = \left( {\begin{array}{*{20}{c}}
  1&1& \cdots &1 \\
  1&2& \ddots & \vdots  \\
   \vdots & \ddots & \ddots &1 \\
  1& \ldots &1&n
\end{array}} \right);b = \left( {\begin{array}{*{20}{c}}
  n \\
  {n + 1} \\
   \vdots  \\
  {2n - 1}
\end{array}} \right).\]
The specified system has a~unique solution~$x = (1, \ldots ,1)$ and has the diagonal dominance property for any~$n \geqslant 2$. We investigated the speedup of the BSF\nobreakdash-\hspace{0pt}Jacobi parallel algorithm by varying the number~$K$ of working nodes.

\begin{table}[t]
\caption{Cost parameters for BSF\nobreakdash-\hspace{0pt}Jacobi parallel algorithm (seconds).}
\centering
\begin{tabular}{l c c c c}
  \hline
  $\textbf{n}$       & \textbf{1\,500}  & \textbf{5\,000}  & \textbf{10\,000} & \textbf{16\,000} \\
  \hline
  $t_c$     & 7.20E-5 & 1.06E-3 & 2.17E-3 & 2.95E-3 \\
  $t_p$     & 5.01E-6 & 1.72E-5 & 3.70E-5 & 5.61E-5 \\
  $t_a$     & 1.89E-6 & 5.27E-6 & 9.31E-6 & 2.10E-5 \\
  $t_{Map}$ & 6.23E-3 & 9.28E-2 & 3.73E-1 & 7.73E-1 \\
  \hline
  $\frac{comp}{comm}$ & 126 & 113 & 215 & 376 \\
  \hline
\end{tabular}\label{Table2}
\end{table}

\begin{figure}[t]
\begin{minipage}[h]{0.49\linewidth}
\centering
\includegraphics[scale=0.35]{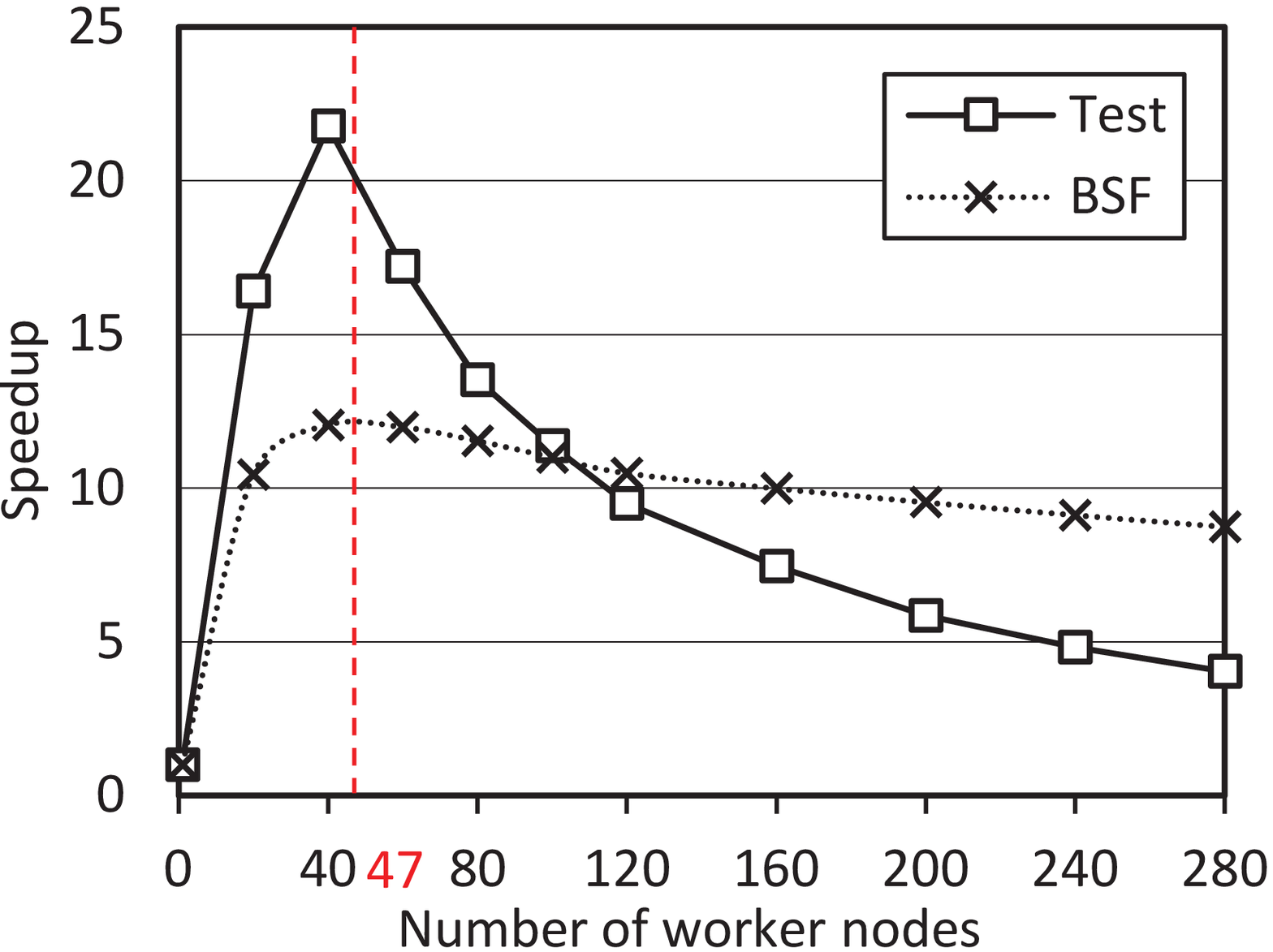}\\
(a) $n=1\,500$
\end{minipage}
\hfill
\begin{minipage}[h]{0.49\linewidth}
\centering
\includegraphics[scale=0.35]{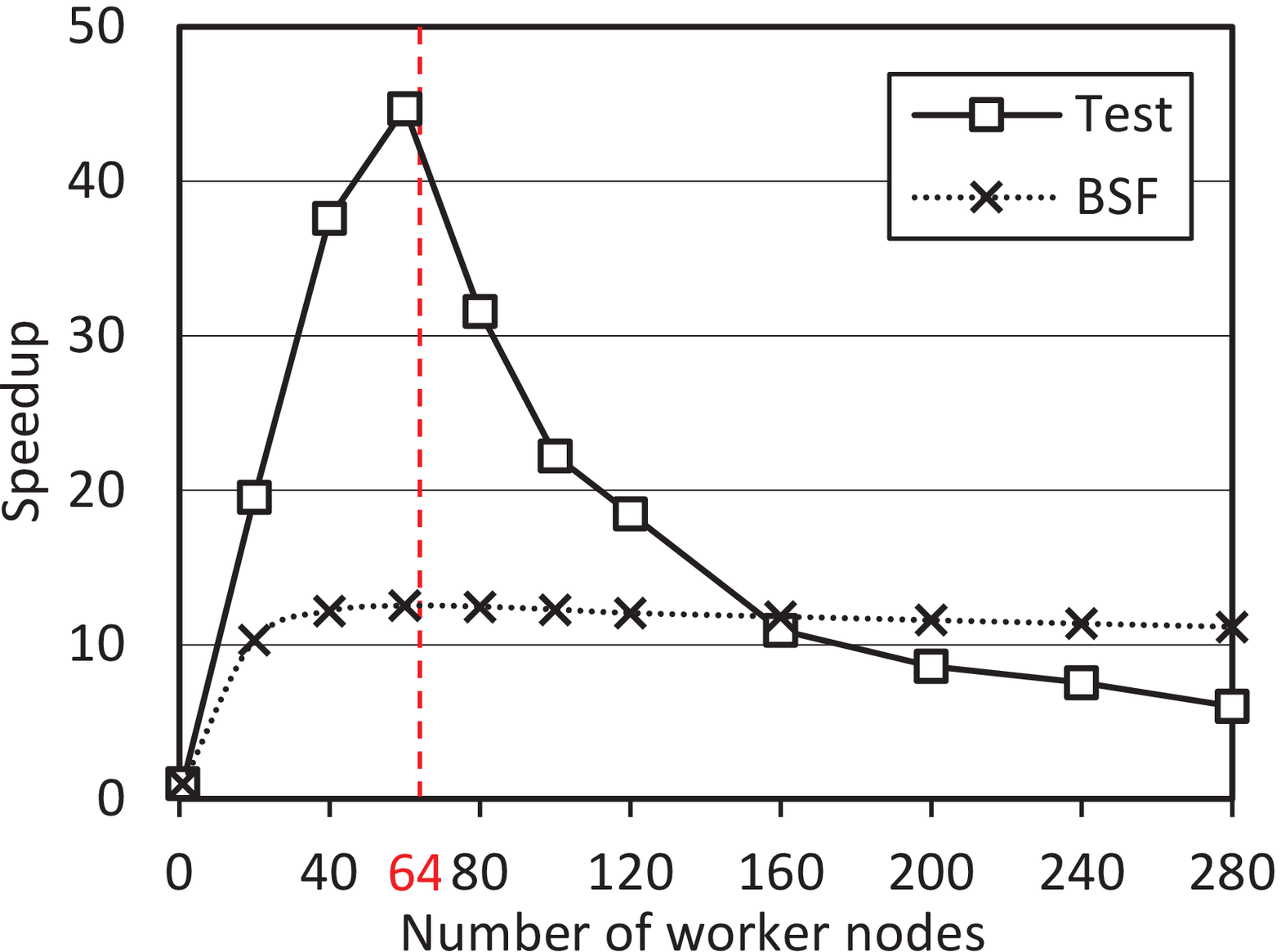}\\
(b) $n=5\,000$
\end{minipage}
\vfill
\bigskip
\begin{minipage}[h]{0.49\linewidth}
\centering
\includegraphics[scale=0.35]{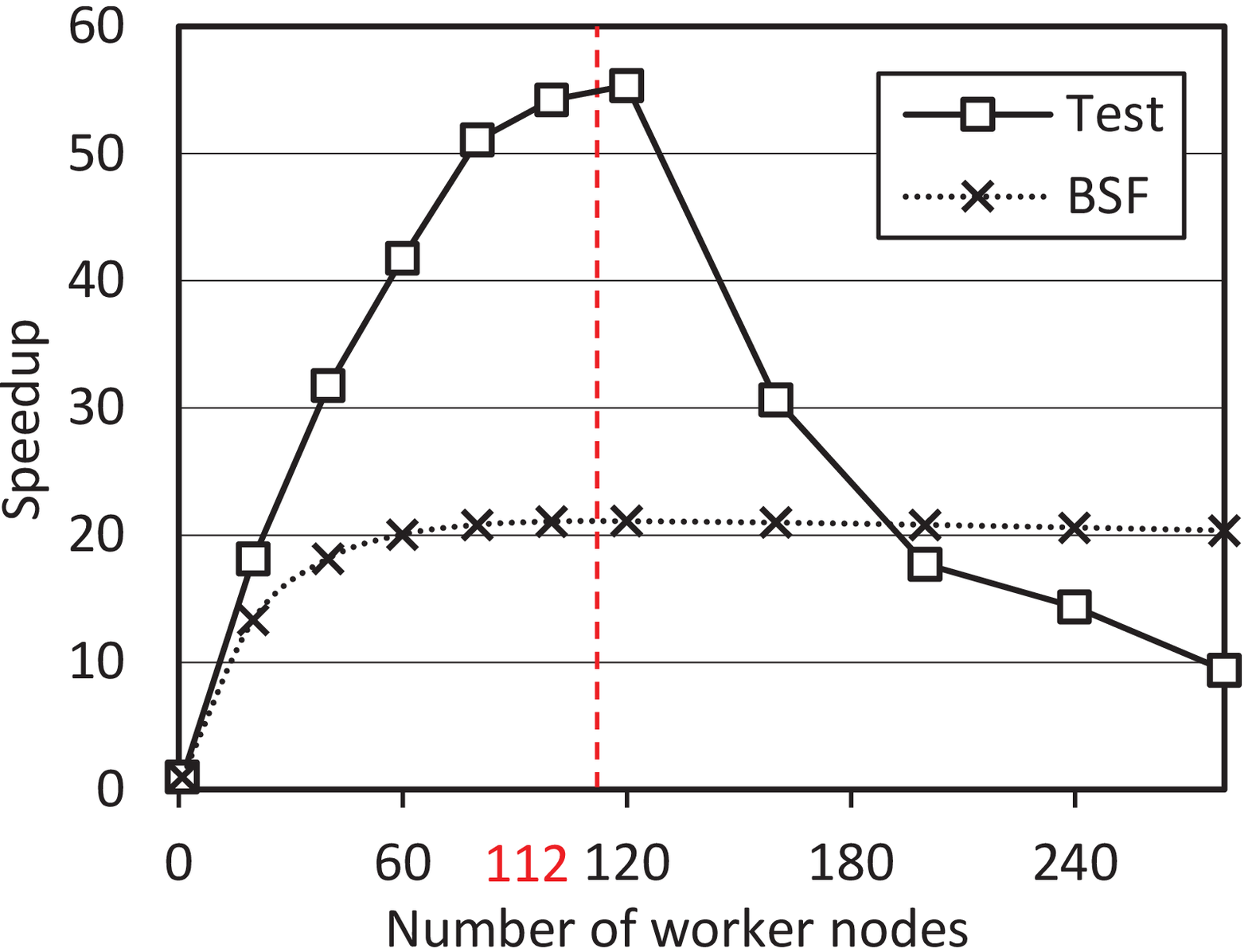}\\
(c) $n=10\,000$
\end{minipage}
\hfill
\begin{minipage}[h]{0.49\linewidth}
\centering
\includegraphics[scale=0.35]{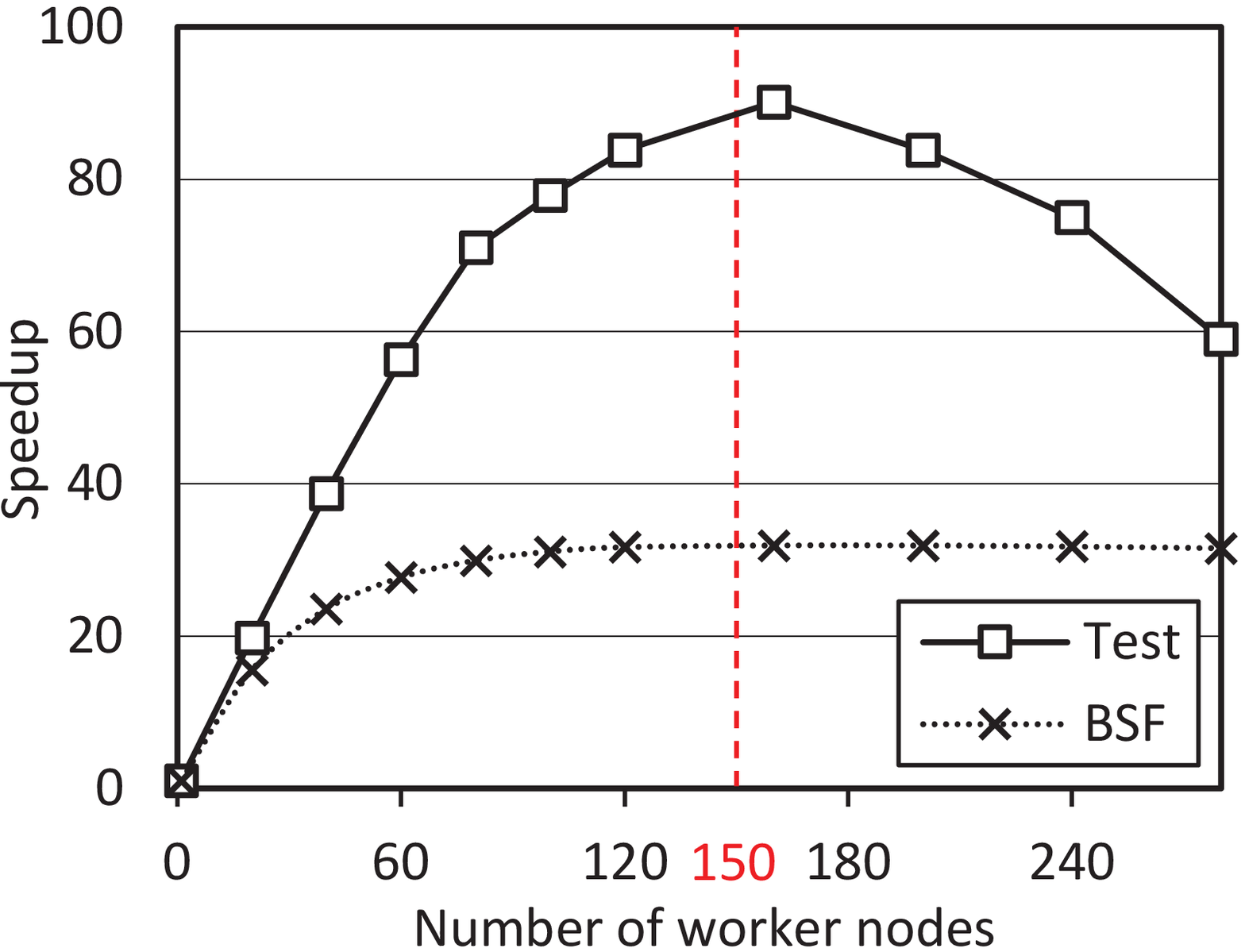}\\
(c) $n=16\,000$
\end{minipage}
\caption{BSF-Jacobi parallel algorithm speedup obtained empirically and theoretically.}
\label{Fig6}
\end{figure}

The speedup~$a_{test}(K)$ was calculated by the equation \[a_{test}(K) = \frac{{{T_1}}}{{{T_K}}},\] where~${T_1}$ is the execution time on configuration with one master node and one worker node, and~${T_K}$ is the execution time on configuration with one master node and~$K$ worker nodes. The computations were performed for dimensions $n = 1\,500$, $n = 5\,000$, $n = 10\,000$ and $n = 16\,000$. The results are presented in Fig.~\ref{Fig6} (the solid curves marked with squares). In the same diagrams, we plotted the curves of speedup calculated analytically by equation~\eqref{equation:09} (the dotted curves marked with crosses). The diagrams in Fig.~\ref{Fig6} also include the values of scalability boundaries obtained using equation~\eqref{equation:14}. This values are flagged with vertical dotted lines. To calculate these analytical estimations, we experimentally determined the values of cost parameters using a configuration with one master and one worker. These values are shown in Table~\ref{Table2}. In all cases the latency $L$ (time of transferring one-byte message node-to-node) was~1.5E-5 sec. We also added the cost ratio between computations  and communications to the table. Here \mbox{$comp=t_{Map}+(n-1)t_a+t_p$} and \mbox{$comm=t_c$}. The cost of computation significantly exceeds the cost of communications in the BSF\nobreakdash-\hspace{0pt}Jacobi algorithm, and this gap tends to increase with the size of the problem. The fact that the value of \mbox{$\tfrac{{comp}}{{comm}}$} for \mbox{$n=1\,500$} is greater than the value for \mbox{$n=5\,000$} is explained as follows. When the cost of a single data exchange becomes comparable to the latency, the latency begins to significantly affect the result.

Let us define the prediction error for the problem of dimension $n$ as follows:
\begin{equation}\label{equation:26}
Error(n) = \frac{|K_{test}(n)-K_{BSF}(n)|}{\max \left(K_{test}(n),K_{BSF}(n)\right)},
\end{equation}
where $K_{test}(n)$ is the speedup boundary obtained experimentally, and $K_{BSF}(n)$ is the speedup boundary obtained analytically by~\eqref{equation:14}. For the Jacobi-BSF parallel algorithm, we obtain the results shown in Table~\ref{Table3}. In all cases, the prediction error does not exceed 15\% and tends to decrease with increasing the problem size.

\begin{table}[t]
\caption{Prediction errors for BSF\nobreakdash-\hspace{0pt}Jacobi parallel algorithm.}
\centering
\begin{tabular}{l c c c c}
  \hline
  $\textbf{n}$ & \textbf{1\,500}  & \textbf{5\,000}  & \textbf{10\,000} & \textbf{16\,000}\\
  \hline
  \textbf{$K_{BSF}$}    & 47      & 64      & 112 & 150\\
  \textbf{$K_{test}$}   & 40      & 60      & 120 & 160\\
  \hline
  \textbf{$Error$}      & \textbf{0.15}    & \textbf{0.06}    & \textbf{0.07}& \textbf{0.06}\\
  \hline
\end{tabular}\label{Table3}
\end{table}

The second series of experiments, the results of which we want to present in this article, relate to a~simplified \emph{n-body problem}~\cite{32} that describes how a~small body will move under the influence of gravitational forces among~$n$ large motionless bodies. Let us give a~brief description of this problem. Let~$\mathbb{Y} \subset {\mathbb{R}^3}$ be a~finite set of points representing motionless bodies of large mass. We will denote these points by~${Y_1}, \ldots ,{Y_n}$, and their masses by~${m_1}, \ldots ,{m_n}$, where~\mbox{$n = \left| \mathbb{Y} \right|$}. Let the point $X$ represent a~body~$x$ of low mass~${m_x}$ moving relative to motionless large mass bodies~${Y_1}, \ldots ,{Y_n}$. We assume that no forces other than gravity act on the body~$x$. We know the initial position~${X^{({t_0})}} \in {\mathbb{R}^3}$ and the velocity vector~\mbox{${V^{({t_0})}} \in {\mathbb{R}^3}$} of the body~$x$ at the instant of time~${t_0}$. The problem is to predict the subsequent motion of the body~$x$ using Newton's laws of motion and Newton's law of universal gravitation. To accomplish this, we will sequentially calculate the following positions of the body~$x$ using a time slot~$\Delta t$:
\begin{equation}\label{equation:27}{X^{({t_0})}},{X^{({t_0} + \Delta t)}},{X^{({t_0} + 2\Delta t)}},{X^{({t_0} + 3\Delta t)}}, \ldots \end{equation}

According to the law of universal gravitation, the gravitational force~${F_i}$ of the mass point~${Y_i}$ acting on the body~$x$ can be calculated using the equation
\begin{equation}\label{equation:28}{F_i} = G\frac{{{m_i}{m_x}}}{{{{\left\| {{Y_i} - X} \right\|}^2}}}({Y_i} - X),\end{equation}
where~$X$ represents the current coordinates of the body~$x$. According to Newton's second law of motion, the acceleration~${\alpha _i}$ of the body~$x$ produced by the force~${F_i}$ can be calculated using the equation
\begin{equation}\label{equation:29}{\alpha _i} = \frac{{{F_i}}}{{{m_x}}}.\end{equation}
The acceleration produced by the all forces~${F_1}, \ldots ,{F_n}$ is calculated by the equation
\begin{equation}\label{equation:30}\alpha  = \sum\limits_{i = 1}^n {{\alpha _i}} .\end{equation}
Consequently, the velocity vectors required to calculate (approximately) the elements of the sequence~\eqref{equation:27} can be calculated using the following iterative equation
\begin{equation}\label{equation:31}{V^{(t + \Delta t)}} = {V^{(t)}} + {\alpha ^{(t + \Delta t)}}\Delta t,\end{equation}
where
\begin{equation}\label{equation:32}{\alpha ^{(t + \Delta t)}} = \sum\limits_{i = 1}^n {G\frac{{{m_i}}}{{{{\left\| {{Y_i} - {X^{(t)}}} \right\|}^2}}}\left( {{Y_i} - {X^{(t)}}} \right)} .\end{equation}
Using~\eqref{equation:31}, we obtain
\begin{equation}\label{equation:33}{X^{(t + \Delta t)}} \approx {X^{(t)}} + {V^{(t + \Delta t)}}\Delta t.\end{equation}

\begin{algorithm}[t]
\caption{BSF-Gravity algorithm.}\label{alg5}
\begin{algorithmic}[1]
\State \textbf{input} $A,X_0,V_0,t_0,T$
\State $t := t_0$; $X := X_0$; $V := V_0$;
\State \label{alg5:loop_begin}$B := Map(f_X,A)$
\State $\alpha := Reduce(\vec +,B)$
\State $\Delta t := Delta\_t(V,\alpha)$
\State $V:=V\vec +\alpha\Delta t$
\State $X:=X\vec +V\Delta t$
\State $t := t+\Delta t$
\State \textbf{if} $t<T$ \textbf{goto} \ref{alg5:loop_begin}
\State \textbf{output} $X$
\State \textbf{stop}
\end{algorithmic}
\end{algorithm}

\begin{algorithm}[t]
\caption{BSF-Gravity parallel algorithm.}\label{alg6}
\begin{multicols}{2}
\begin{center}
\textbf{Master} \\
\textbf{\emph{j}th Worker (j=1,\dots,K)}
\end{center}
\end{multicols}
\algrule
\begin{multicols}{2}
\begin{algorithmic}[1]
\State \textbf{input} $X_0,V_0,t_0,T$
\State $t := t_0$; $X := X_0$; $V := V_0$;
\State\label{alg6:Master_loop_begin}$SendToAllWorkers(X)$
\State
\State
\State $RecvFromWorkers\left(\alpha_1,\ldots,\alpha_K\right)$
\State $\alpha:= Reduce\left(\vec+,[\alpha_1,\dots,\alpha_K]\right)$
\State $\Delta t := Delta\_t(V,\alpha)$
\State $V:=V\vec +\alpha\Delta t$
\State $X:=X\vec +V\Delta t$
\State $t := t+\Delta t$
\State $exit := t<T$
\State $SendToAllWorkers(exit)$
\State \textbf{if not} $exit$ \textbf{goto} \ref{alg6:Master_loop_begin}
\State \textbf{output} $X$
\State \textbf{stop}
\end{algorithmic}
\begin{algorithmic}[1]
\State \textbf{input} $A_j$
\State
\State \label{alg6:Worker_loop_begin}$RecvFromMaster\left(X\right)$
\State $B_j := Map(f_X,A_j)$
\State $\alpha_j := Reduce(\vec+,B_j)$
\State $SendToMaster(\alpha_j)$
\State
\State
\State
\State
\State
\State
\State $RecvFromMaster(exit)$
\State \textbf{if not} $exit$ \textbf{goto} \ref{alg6:Worker_loop_begin}
\State
\State \textbf{stop}
\end{algorithmic}
\end{multicols}
\end{algorithm}

\begin{figure}[t]
\begin{minipage}[h]{0.49\linewidth}
\centering
\includegraphics[scale=0.35]{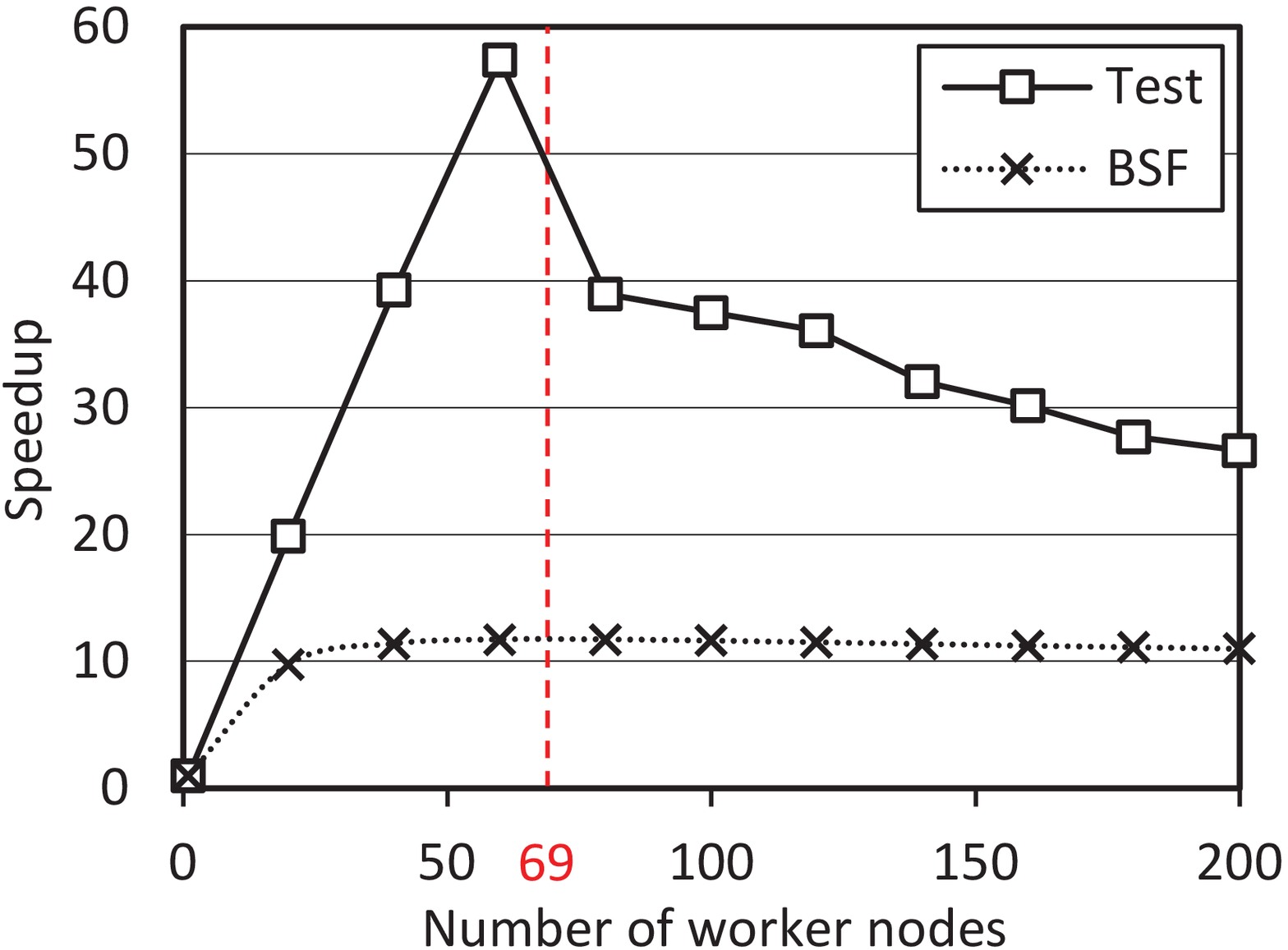}\\
(a) $n=1\,500$
\end{minipage}
\hfill
\begin{minipage}[h]{0.49\linewidth}
\centering
\includegraphics[scale=0.35]{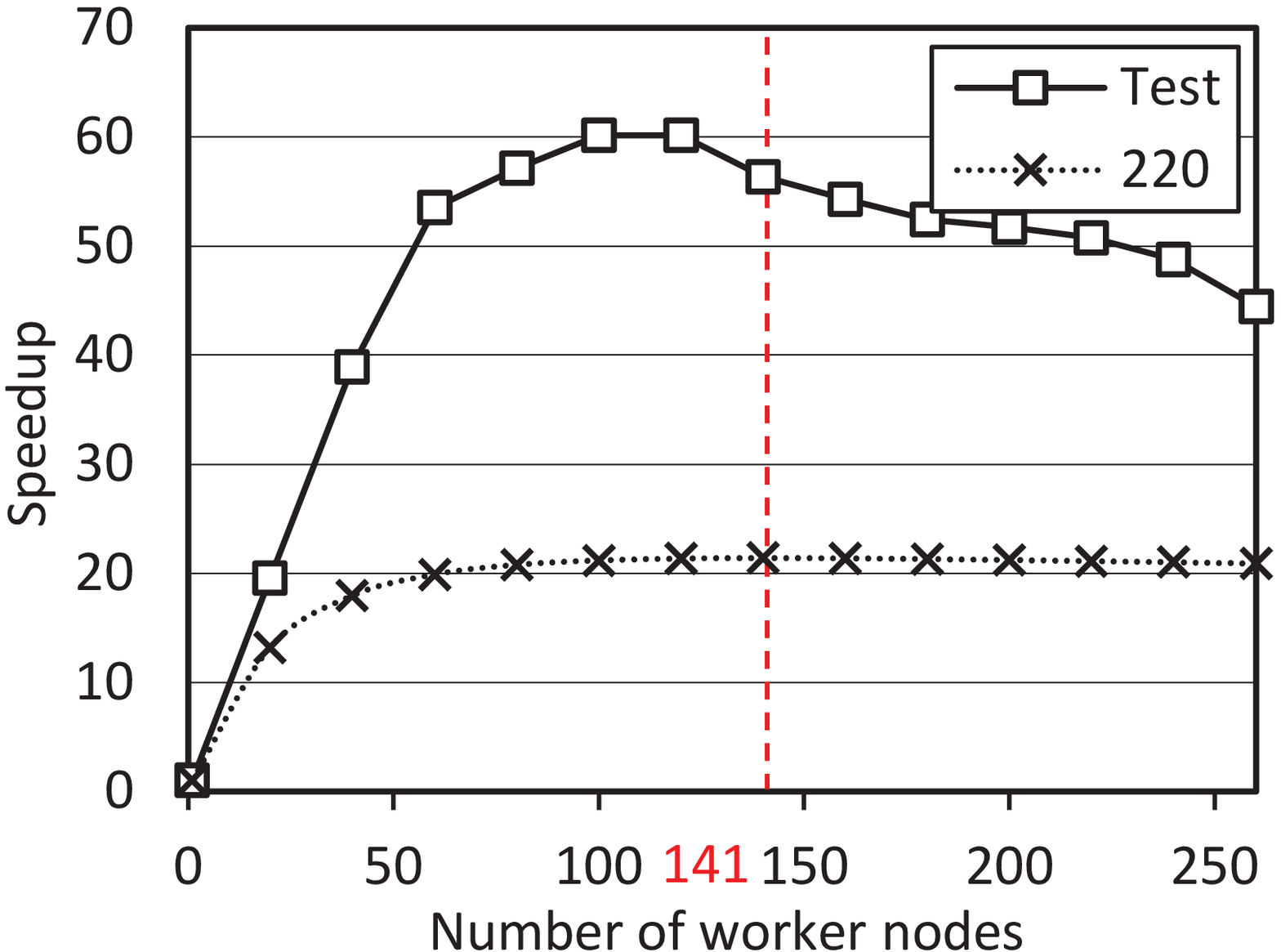}\\
(b) $n=5\,000$
\end{minipage}
\vfill
\bigskip
\begin{minipage}[h]{0.49\linewidth}
\centering
\includegraphics[scale=0.35]{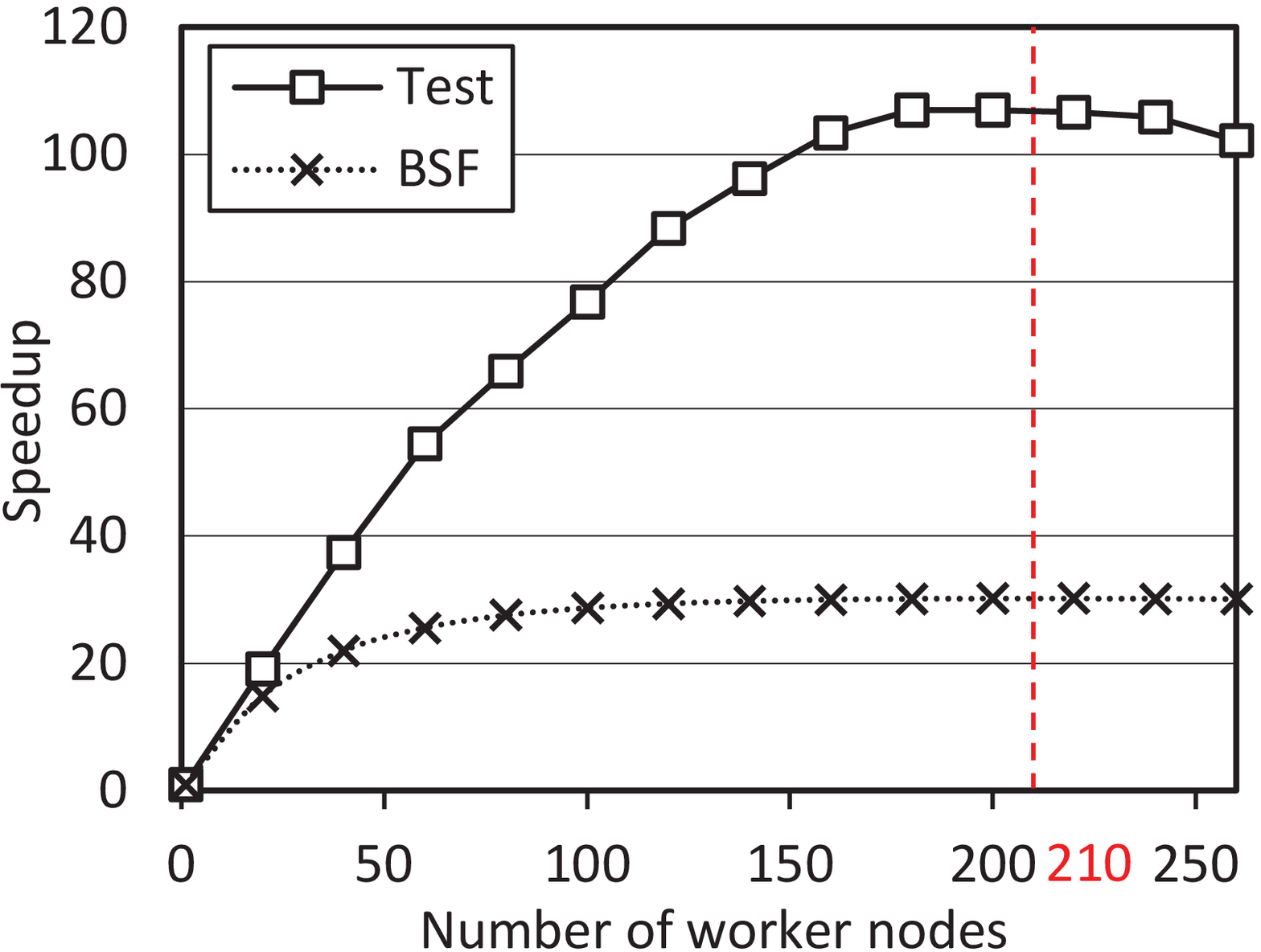}\\
(c) $n=10\,000$
\end{minipage}
\hfill
\begin{minipage}[h]{0.49\linewidth}
\centering
\includegraphics[scale=0.35]{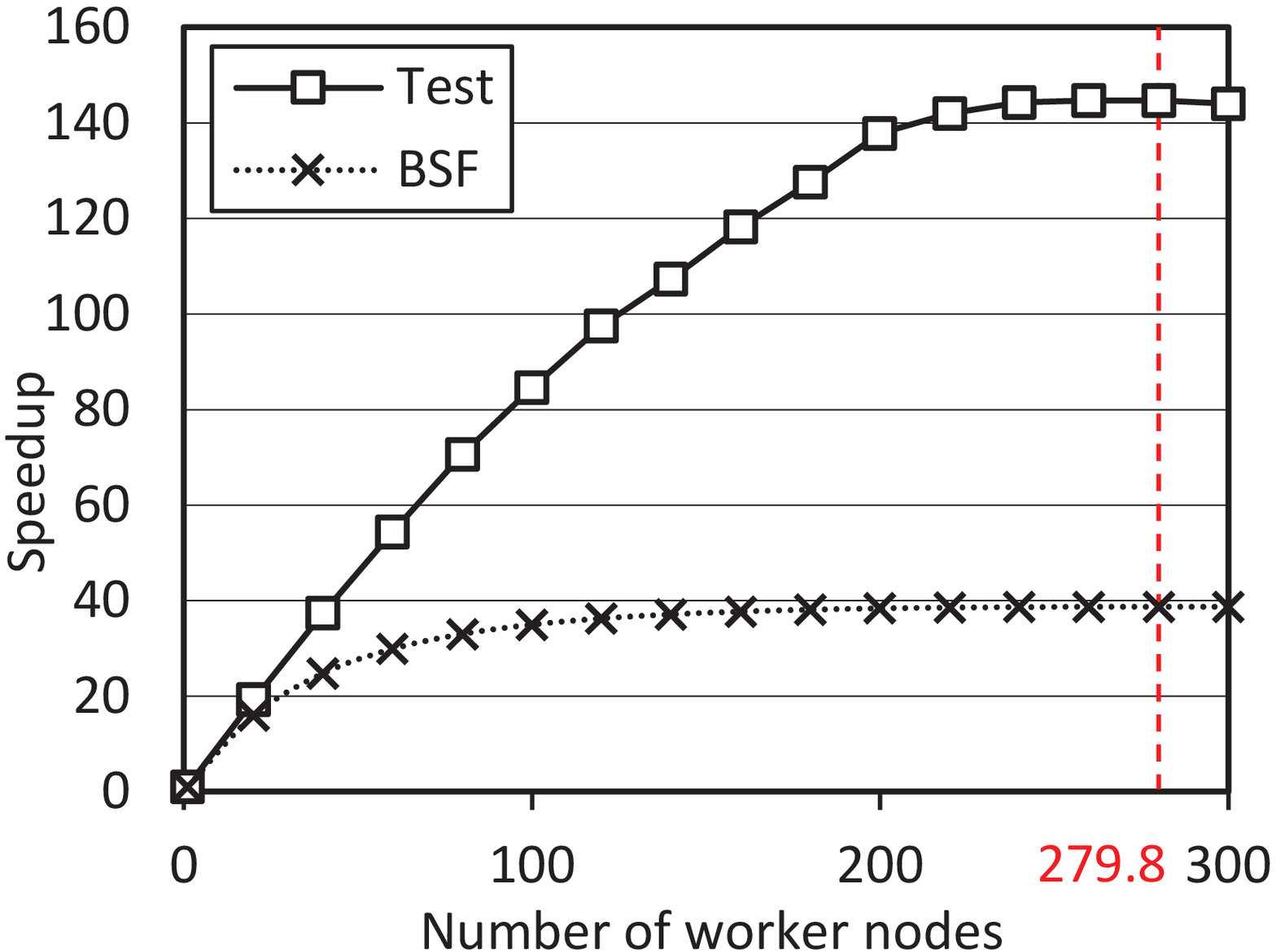}\\
(c) $n=16\,000$
\end{minipage}
\caption{BSF-Gravity parallel algorithm speedup obtained empirically and theoretically.}
\label{Fig7}
\end{figure}

Using the generic BSF\nobreakdash-\hspace{0pt}algorithm template, we obtain Algorithm~\ref{alg5} named the \emph{BSF\nobreakdash-\hspace{0pt}Gravity} that presents an implementation of the simplified $n$-body problem. In the context of this problem, $A$ is the list of pairs~$({Y_i},{m_i})$ that specify the coordinates and mass of the~$i$\nobreakdash-\hspace{0pt}th motionless large body:
\begin{equation}\label{equation:34}A = [({Y_1},{m_1}), \ldots ,({Y_n},{m_n})];\end{equation}
$T$ is the instant of time that culminates the calculation of the trajectory of the body~$x$; $f_X$ is the parameterized function defined by the equation
\begin{equation}\label{equation:35}f_X({Y_i},{m_i}) = G\frac{{{m_i}}}{{{{| Y_i - X |}^2}}}({Y_i} - X)\end{equation}
that the higher-order function \emph{Map} takes as a~parameter to apply to items of the list~$A$; and~$\vec  + $~denotes the operation of vector addition in~${\mathbb{R}^3}$. The user function $Delta\_t(V,\alpha)$ calculates the time slot $\Delta t$ depending on the current velocity~$V$ and acceleration~$\alpha$.
The parallel implementation of the BSF\nobreakdash-\hspace{0pt}Gravity algorithm is obtained automatically by using the generic BSF-algorithm parallelization template (Algorithm~\ref{alg2}). This implementation is presented in Algorithm~\ref{alg6}.

Let us evaluate Algorithm~\ref{alg6}. Assume that the user function $Delta\_t(V,\alpha)$ is defined as follows:
\[Delta\_t(V,\alpha ) = \frac{\eta }{{{\left\| V \right\|}^2 \cdot {{\left\| \alpha  \right\|}^4}}},\]
where $\eta$ is a positive constant. Calculating this function takes 13 arithmetic operations. The analysis of Algorithm~\ref{alg6} gives us the following estimations: $t_c=6\tau _{tr}+2L$, $t_{Map}=17n\tau _{op}$, $t_a=3\tau _{op}$ and $l=n$. Substituting these quantities in~\eqref{equation:14}, we obtain the following scalability boundary for the BSF\nobreakdash-\hspace{0pt}Gravity parallel algorithm:
\begin{equation}\label{equation:36}
K_{BSF - Gravity} = \frac{1}{2}\sqrt {{{\left( {\frac{{6{\tau _{tr}} + 2L}}{{3{\tau _{op}}\ln 2}}} \right)}^2} + \frac{{29}}{3}n}  - \frac{{6{\tau _{tr}} + 2L}}{{3{\tau _{op}}\ln 2}} .
\end{equation}
For~$n \to \infty $, equation~\eqref{equation:36} asymptotically tends to the following estimation:
\begin{equation}\label{equation:37}{K_{BSF-Gravity}} \approx O(\sqrt{n}),\end{equation}
where~$n$ is the number of motionless large bodies.

We implemented the BSF\nobreakdash-\hspace{0pt}Gravity parallel algorithm using the BSF\nobreakdash-\hspace{0pt}skeleton. The source code of this implementation is freely available on GitHub, at \url{https://github.com/leonid-sokolinsky/BSF-gravity}. Using this implementation, we conducted the computational experiments on the ``Tornado SUSU'' computing cluster. The computations were performed for the following numbers of large bodies: $n = 300$, $n = 600$, $n = 900$ and $n = 1\,200$. A comparison of the results obtained empirically and theoretically (by using equation~\eqref{equation:09}) is shown in Fig.~\ref{Fig7}. To plot the curve tagged as ``BSF'', we experimentally determined the following cost parameters that are independent of~$n$: $t_c=5\cdot 10^{-5}$, $t_p=9.5\cdot 10^{-7}$, $t_a=4.7\cdot 10^{-9}$ and $L=1.5\cdot 10^{-5}$ (in seconds). For the parameter $t_{Map}$ that depends on $n$, the following values in seconds were obtained: $3.6\cdot 10^{-3},7.46\cdot 10^{-3},1.12\cdot 10^{-2},1.5\cdot 10^{-2}$ for $n$ equal to $300, 600, 900$ and $1200$, respectively. The cost ratio between computations and communications varied from $100$ at $n=300$ to $411$ at $n=1200$. The error values calculated using equation~\eqref{equation:26} are shown in table~\ref{table:4}. In all cases, the error does not exceed 13\% and tends to decrease with increasing the problem size.

\begin{table}[t]
\caption{Prediction errors for BSF\nobreakdash-\hspace{0pt}Gravity parallel algorithm.}
\centering
\begin{tabular}{l c c c c}
  \hline
  $\textbf{n}$ & \textbf{300}  & \textbf{600}  & \textbf{900} & \textbf{12\,000}\\
  \hline
  \textbf{$K_{BSF}$}    & 69      & 141      & 210 & 279.1\\
  \textbf{$K_{test}$}   & 60      & 140      & 200 & 280\\
  \hline
  \textbf{$Error$}      & \textbf{0.13}    & \textbf{0.01}    & \textbf{0.05}& \textbf{3.6E-4}\\
  \hline
\end{tabular}\label{table:4}
\end{table}

Another example of using the BSF model can be found in~\cite{33}, where we investigate an iterative numerical method for solving nonstationary systems of linear inequalities. In this article, we also compare the speedup curves constructed analytically using the BSF model and the speedup curves obtained by conducting experiments on the real cluster computing system. All conducted experiments confirm the adequacy of the BSF model.

\section{Discussion}\label{Section:Discussion}

In this section, we discuss the strengths and weaknesses of the BSF model and answer the following questions.
\begin{enumerate}[1.]
\item Is it possible to use the BSF model for the algorithms that process sets?
\item Can we apply the BSF model to an algorithm that uses only the \emph{Map} function without the \emph{Reduce} function?
\item Can we apply the BSF model to a~numerical algorithm that is not iterative?
\item What is the difference between MapReduce and the BSF model?
\item Does the BSF model admit a~configuration of the BSF\nobreakdash-\hspace{0pt}computer with two or more master nodes?
\item Does the BSF model take into account the multicore structure of the processor node?
\item Is the BSF model the best predictor of the execution time of a~parallel algorithm on a~target multiprocessor system?
\end{enumerate}

Let us start by discussing the advantages of the BSF model. The main contribution of the BSF model and this article is equation~\eqref{equation:14} that allows us to estimate the scalability boundary of a~parallel program at an early stage of its design. No other known model yields such an equation. In addition, the BSF model is easy to use when designing and analyzing parallel algorithms and programs. Based on the BSF model, we constructed a~compilable algorithmic skeleton using the MPI library that allows quick creation of a~syntactically valid BSF program. However, to acquire this result, we introduced a~number of constraints.

First, the BSF model requires the representation of a~numerical method as an algorithm over lists using the higher-order functions \emph{Map} and \emph{Reduce}. How strong is this restriction? We will start by answering the question whether the BSF model can be applied to multisets (sets that allows duplicates)~\cite{34}. Let~$\mathcal{A} = \{ {a_1}, \ldots ,{a_n}\} $ be a~finite multiset. By ordering, we can represent~$\mathcal{A}$ as the list~$A = [{a_1}, \ldots ,{a_n}]$. Let~$B = Map(f,A) = \left[ {f({a_1}), \ldots ,f({a_n})} \right]$. Ignoring ordering, we can transform~$B$ to the multiset $\mathcal{B} = \left\{ {f({a_1}), \ldots ,f({a_n})} \right\}$. Thus, we can apply the BSF model to multisets without any modifications. Applying the BSF model to sets is also possible if the function~$f$ does not generate duplicates. Taking into account the generic nature of the \emph{Map} function, we can conclude that the representation of the numerical method as an algorithm over lists is not a~very strong restriction.

The second question is: Can we apply the BSF model to an algorithm that uses only the \emph{Map} function without the \emph{Reduce} function? The answer is ``yes''. In this case, the parameter~${t_a}$ denoting the time of execution of the operation~$\oplus$ is assumed to be zero. An example of applying the BSF model to an algorithm using only the \emph{Map} function is presented in~\cite{35}.

Next, we limited the scope of the BSF model application to the iterative compute-intensive numerical algorithms. The iterative nature of the algorithm assumes that the iterative process takes much longer than initialization (reading or generating the problem data, allocating the memory, etc.), and we can neglect the cost of the last (there is no corresponding parameter in the BSF cost metric). Consequently, we can use the BSF model if our noniterative algorithm is presented in the form of operations over lists, and the initialization cost is negligible compared to the execution cost of the \emph{Map} and \emph{Reduce} functions. This is the answer to the third question stated above.

The compute-intensive nature of the algorithm assumes that the cost of computations is greater than or comparable to the cost of interprocessor communications and input/output operations. If we have the opposite situation, when the communications and input/output operations significantly exceed the calculations, then, due to property~\eqref{equation:12}, we have~${K_{BSF}} = 1$, and the BSF model becomes inapplicable. This is the main difference between the BSF model and the programming model MapReduce~\cite{36} that is intended for processing and generating big data sets. This is the answer to the 4th question.

The 5th question is: Does the BSF model admit a~configuration of the BSF\nobreakdash-\hspace{0pt}computer with two or more master nodes? The answer is ``no''. All attempts to derive an equation such as~\eqref{equation:14} failed in the case of configurations with two or more master nodes.

The 6th question is: Does the BSF model take into account the multicore structure of the processor node? The answer is also ``no'': the model treats the processor node as a black box that can perform scalar and vector operations at a certain speed. Most of the modern cluster computing systems have the processor nodes that include multicore processors and GPUs. The efficient usage of such systems is impossible without the use of intranode parallelism and vectorization. However, in this situation, how can we determine the adequate values of the cost parameters~${t_{Map}}$ (the time taken by a~single processor node to process the whole list by using the higher-order function \emph{Map}) and~${t_a}$ (the time taken by a~processor node to execute the operation~$\oplus$)? Let us explain the possible solution of this problem by the following example. Let~$\oplus$ be the addition operation of two vectors of dimension 1\,000\,000. We write a~program that computes the sum of 1\,000\,000 such vectors using all resources of the intranode parallelism and vectorization. Then, we run this program on a~single processor node of the target system and measure the execution time (excluding initialization). After that, we divide the resulting time by 1\,000\,000 and obtain the value of the parameter~${t_a}$. In the same way, we can obtain the value of the parameter~${t_{Map}}$. This is a~quite rough method for obtaining the values of these parameters. However, the main goal of the BSF model is to predict the scalability boundary being the maximum number of processor nodes to which the speedup increases. If a~more accurate prediction of the execution time of a~parallel program on a~computing cluster with multicore nodes is needed, then other parallel computation models should be used (see, for example,~\cite{37,38}). Therefore, the BSF model is not the best predictor of the execution time of a~parallel algorithm on a~target multiprocessor system. This is the answer to the last question of our discussion.

\section{Conclusions}\label{Section:Conclusions}

In this paper, we presented a~novel parallel computation model named bulk synchronous farm (BSF) that is an extension of the BSP (bulk-synchronous parallel) model. The main advantage of the proposed model is that it allows to estimate the scalability of a parallel algorithm before its implementation. Another important feature of the BSF model is the representation of problem data in the form of lists that greatly simplifies the logic of building applications. To develop the BSF model, we used a~new approach that restricts not only the class of multiprocessor architectures but also the type of algorithms admitted by the model. The application scope of the BSF model is algorithms of the compute-intensive iterative type performed on cluster computing systems. In the BSF model, the processor nodes are organized by using the master/slave paradigm. The BSF model uses the Map/Reduce scheme to parallelize applications. We constructed a~cost metric of the BSF model. This metric requires the representation of a~numerical method as an algorithm over lists using the higher-order functions \emph{Map} and \emph{Reduce}. Using this metric, we derived equation~\eqref{equation:14} that allowed us to estimate the scalability boundary of a~parallel algorithm before its software implementation. No other known model yields such an equation. Based on the BSF model, we constructed a~compilable algorithmic skeleton using the MPI library that allows the quick creation of a~syntactically valid BSF\nobreakdash-\hspace{0pt}program. Using this skeleton, we implemented several iterative algorithms and conducted large-scale computational experiments on a~real cluster computing system. In all cases, experiments have shown that the scalability boundary of the algorithm obtained analytically using the BSF cost metric is very close to the scalability boundary obtained experimentally. Our experience with the BSF model has shown that this model is accurate and easy to use when designing and analyzing parallel algorithms and programs.

\section{Declaration of competing interests}

The author declare that he has no known competing financial interests or personal relationships that could have appeared to influence the work reported in this paper.

\section{Acknowledgments}

This work was supported by the Russian Foundation for Basic Research [project No.~20-07-00092-a] and the Ministry of Science and Higher Education of the Russian Federation [gov. order FENU-2020-0022].


\end{document}